\documentclass[10pt,a4paper,twoside]{aa}
		
\usepackage{graphicx}

\begin{document}

\title{\textit{XMM-Newton} observation of Kepler's supernova remnant}

\titlerunning{\textit{XMM-Newton} observation of Kepler's SNR}

\author{\textsc{G.~Cassam-Chena\"{i}}\inst{1} 
\and \textsc{A.~Decourchelle}\inst{1} 
\and \textsc{J.~Ballet}\inst{1}
\and \textsc{U.~Hwang}\inst{2} 
\and \textsc{J.P.~Hughes}\inst{3} 
\and \textsc{R.~Petre}\inst{4}}

\authorrunning{\textsc{G.~Cassam-Chena\"{i}} et al.}

\offprints{Gamil \textsc{Cassam-Chena\"{i}}, \\
\email{gcc@discovery.saclay.cea.fr}}

\institute{Service d'Astrophysique, CEA Saclay, 91191 Gif-sur-Yvette Cedex, France
\and Department of Astronomy, University of Maryland, College Park, MD 20742
\and Department of Physics and Astronomy, Rutgers University, 136 Frelinghuysen Road, Piscataway, NJ 08854-8109
\and Laboratory for High Energy Astrophysics, Goddard Space Flight Center, Greenbelt, MD 20771}

\date{}

\abstract{We present the first results coming from the observation of Kepler's supernova
remnant obtained with the EPIC instruments on board the \textit{XMM-Newton} satellite. 
We focus on the images and radial profiles of the emission lines (Si K, Fe L, Fe K) and of the high energy continuum.
Chiefly, the Fe L and Si K emission-line images are generally consistent with each other and the
radial profiles show that the Si K emission extends to a larger radius than the Fe L emission 
(distinctly in the southern part of the remnant).
Therefore, in contrast to Cas A, no inversion of the Si- and Fe-rich ejecta layers is observed in Kepler.
Moreover, the Fe K emission peaks at a smaller radius than the Fe L emission, which implies that the temperature increases inwards in the ejecta.
The 4-6 keV high energy continuum map shows the same distribution as the asymmetric emission-line images
except in the southeast where there is a strong additional emission.
A two color image of the 4-6 keV and 8-10 keV high energy continuum
illustrates that the hardness variations of the continuum are weak all along the remnant except in a few knots.
The asymmetry in the Fe K emission-line is not associated with any asymmetry in the Fe K equivalent width map.
The Si K maps lead to the same conclusions.
Hence, abundance variations do not cause the north-south brightness asymmetry. The strong emission
in the north may be due to overdensities in the circumstellar medium. 
In the southeastern region of the remnant, the lines have a very low equivalent width and the X-ray emission is largely nonthermal.
}

\maketitle

\section{Introduction}

The remnants of recent stellar explosions give out a large part of their
luminosity in X-rays. The interaction of the high velocity supernova material (ejecta)
with the ambient medium gives rise to a forward shock moving into the ambient medium
and a reverse shock propagating back into the ejecta (\cite{mckee}).
Both shocks compress and heat the matter. 
Thus the X-ray emission arises from two adjacent media, different
in composition, density and temperature: the shocked ejecta and the
shocked ambient medium.

At an age of nearly 400 years, Kepler is a young supernova remnant (SNR) whose X-ray emission is
still dominated by a thermal spectrum characteristic of shocked ejecta. 
It is about $200\arcsec$ in angular size. At a distance of about $4.8\pm 1.4$ kpc (\cite{reynoso}), this
corresponds to an angular radius of about $2.3\pm 0.7$ pc. 
This shell-type remnant shows a strongly asymmetric brightness distribution in X-rays.
Well correlated optical and infrared observations also exhibit an asymmetric emission, 
with the remnant being notably brighter in the north and northwest with
some structures near the center of the remnant (Bandiera \& van den Bergh 1991; Douvion et al. 2001).
The optical knots in Kepler's SNR, which have a low expansion velocity (Bandiera \& van den Bergh 1991), a high
density and nitrogen overabundances (Dennefeld 1982; Blair et al. 1991), are attributed to a circumstellar origin.
The measurement in a Balmer-dominated knot of a large [NII] line width favors a nonthermal broadening in a cosmic-ray precursor (Sollerman et al. 2003).
As for the radio emission, it has the same shell-like morphology as the X-rays
with the same north-south brightness asymmetry (Matsui et al. 1984; Dickel et al. 1988).
Otherwise, Kepler's SNR's type is not clearly established.
On one hand, its position above the Galactic plane ($\approx 600$ pc) and
its geometrical similarities with Tycho's SNR favor a type Ia progenitor (\cite{smith}) whereas, 
on the other hand, optical and infrared observations
which reveal the presence of circumstellar material suggest a type II progenitor.
The light-curve does not allow to distinguish between a type I (\cite{baade}) 
and a type II-L (\cite{doggett}) classification.

Previous X-ray studies have been made on images and spectra, but always separately.
The best images of Kepler's SNR were obtained with the \textit{Einstein} HRI (White \& Long 1983; Matsui et al. 1984)
and \textit{ROSAT} HRI (\cite{hughes2}). The upper limit of the energy range did not exceed 4 keV and no spectro-imaging was available.
Spectral studies were more numerous. The SSS on board the \textit{Einstein} satellite revealed
line emission from the He-like species of Si, S and Ar (\cite{becker}). The 1 to 10 keV spectra obtained
with the \textit{EXOSAT} and \textit{Ginga} observatories have shown a strong iron K$\alpha$ emission line at nearly 6.5 keV (\cite{smith};
\cite{hatsukade}). \textit{ASCA} data confirmed these facts with a better energy resolution (\cite{kinugasa}).
From these spectra, several models have been tested coupling non-equilibrium conditions with the hydrodynamic 
evolution of the SNR (Hughes \& Helfand 1985; Decourchelle \& Ballet 1994; Rothenflug et al. 1994; Decourchelle et al. 2000).

A more specific model (\cite{borkowski}) was applied following Bandiera's idea (\cite{bandiera1})
in which the progenitor was a massive runaway mass-losing star ejected from the Galactic plane.
A specific review on Kepler's SNR can be found in Decourchelle (2000).

The new generation satellites \textit{XMM-Newton} and \textit{Chandra}
mark a new era in observing capabilities. Indeed, the onboard technology is made of
CCD detectors associated with telescopes allowing combined good spectral and
spatial resolution with in addition the capability to collect a large number of photons.
Furthermore, both satellites are equipped with slitless gratings giving access to high resolution spectroscopy 
for point or slightly extended sources particularly around 1 keV (around the Fe L complex).
These improvements enable us to analyse thoroughly the spectra of small-scale structures of a SNR as well as
to study maps in a given energy band. This kind of work has been done for several SNRs:
Cas A (Hughes et al. 2000; Hwang et al. 2000; Bleeker et al. 2001; Gotthelf et al. 2001; Willingale et al. 2002, Willingale et al. 2003), 
Tycho (Decourchelle et al. 2001; Hwang et al. 2002), 
the Crab Nebula (Willingale et al. 2001).

Thanks to the EPIC imaging spectrometers of \textit{XMM-Newton} observatory, it is
henceforth possible to carry on studying more precisely Kepler's SNR.
In this paper, we intend to draw up maps of the brightest X-ray emission lines and to lay emphasis on the high energy continuum.
Together with this global approach, we present early results coming from a local spectral study of the forward shock.

\section{Data reduction}\label{data red}

Kepler's SNR was observed on 2001 March 10 using the \textit{XMM-Newton} EPIC and RGS instruments.
The result of the RGS analysis is complicated by the large spatial extent of the source and
will be the purpose of a forthcoming paper.
Here, we focus only on the photons collected by the EPIC instruments~: MOS1, MOS2 and pn.
The thick filter has been used to limit the number of low-energy photons and hence pile-up.
For the pn camera, we have chosen the large window mode in order to limit the out-of-time events.
The obtained data quality is very good (few flares). 
We have used the Science Analysis System (SAS, version 5.3) for data reduction.

We have discovered in the MOS2 data the presence of a few bright pixels in the central CCD which
shift and broaden the spectrum of the whole column (in the readout-out direction) containing these bad pixels. 
They existed before the launch of \textit{XMM-Newton} but were never detected. 
Consequently, we have removed the entire columns RAWX=273, 276, 277 and the column RAWX=383 from RAWY=311 to RAWY=600 in the MOS2 events list.

Because the source is bright, it is necessary to estimate pile-up (\cite{ballet}).
The maximum brightness (in cts/frame/pixel) is equal to 0.0086 for MOS and 0.0048 for pn.
We can estimate the flux loss and the pile-up rate at the maximum brightness\footnote{The result depends on the proportion of X-rays creating a charge pattern with $i$ pixels called $\alpha_i$ in Ballet (1999). Here $\alpha_i = [0.708;0.233;0.021;0.038]$ (from the data, averaged over energy).}.
The results are shown in \mbox{Table \ref{pileup}}.
They show that pile-up is not enough to alter our results, although the true images would show slightly more contrast.

\begin{table}[t]
\centering
\begin{tabular}{lcc}
\hline
           & Flux loss & Pile-up rate \\ \hline \hline
Single Events & $12.7 \%$ & $0.5 \%$ \\
Double Events   & $10.1 \%$ & $7.5 \%$ \\ \hline
\end{tabular}
\caption{Flux loss and pile-up rate in the worst case
corresponding to the 0.0086 cts/frame/pixel of MOS.}
\label{pileup}
\end{table}

\begin{table}[t]
\centering
\begin{tabular}{cccc}
\hline
 & Si K & S K & Fe K \\ \hline \hline
MOS1 & $1.8606\pm{0.0004}$ & $2.4465\pm{0.0010}$ & $6.4655\pm{0.0050}$ \\
MOS2 & $1.8540\pm{0.0004}$ & $2.4374\pm{0.0010}$ & $6.4374\pm{0.0050}$ \\
pn   & $1.8471\pm{0.0004}$ & $2.4304\pm{0.0009}$ & $6.4481\pm{0.0028}$ \\
\textit{ASCA} & $1.854\pm{0.002}$ & $2.445\pm{0.003}$ & $6.46\pm{0.01}$ \\ \hline
\end{tabular}
\caption{Fitted centroid energies (in keV) of three prominent line blends in Kepler's SNR after CTI correction as from the integrated spectrum. The errors are in the range 
$\Delta \chi^2 < 2.7$ (90\% confidence level).}
\label{cti}
\end{table}

To create light curves, spectra and images, we have selected the single, double, triple and 
quadruple events (pattern$\leq$12) for the MOS cameras, and the singles and doubles (pattern$\leq$4) for the pn camera 
(with FLAG$=$0 for all cameras).
The data were cleaned from the few flares using the light curves between 10-12 keV for MOS and 12-14 keV for pn.
Then, the exposure time of the observation amounts to 30 ks for each MOS camera
and to 27 ks for the pn camera.
Although MOS data were corrected for Charge-Transfer
Inefficiency (CTI), a spectral shift depending on energy remains (see Table \ref{cti}).
The fitted centroid energies from the EPIC instruments and the \textit{ASCA} data 
(Kinugasa \& Tsunemi 1999) emphasize that
a single additive or multiplicative factor for each data set cannot be found 
so that the individual lines (e.g., Si K, S K and Fe K)
can be made mutually consistent for the various instruments.
Moreover, since there are no absolute position for the line blends 
due to non-equilibrium effects, 
we cannot re-calibrate these line blends at a reference energy.

\begin{figure}[t]
\centering
\includegraphics[width=6cm,angle=-90]{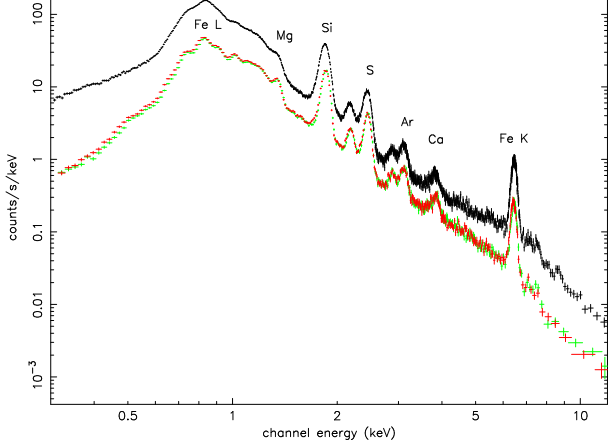}
\caption{The EPIC spectra obtained for the entire remnant (MOS below and pn above).
The energy bands in which we are interested are indicated in Table \ref{bands}.}
\label{spectre_all}
\end{figure}

Kepler's SNR is contained within the central CCD of the MOS cameras and
spread over 4 CCDs of the pn camera resulting in gaps between each CCD. Below 4 keV, we have only used
the MOS data since we have enough statistics to obtain precise images without gaps. 
Above 4 keV, the MOS data have been
combined with the pn's to remedy the lack of statistics at higher energy but introducing gaps in the images.
Note that the spectral analysis has been carried out over 0.3-10 keV and 0.3-12 keV for MOS and pn, respectively.

In this paper, we did not use a local background to be subtracted (in the spectra as well as in the images) since
Kepler's SNR is so bright that the background is contaminated by the wings of the PSF over the entire field of view (FOV).
Thus, we have only subtracted a 400 ks astrophysical and instrumental background (D. Lumb et al. 2002) which
has been renormalized in the 10-12 keV band for MOS and 12-14 keV band for pn over the full FOV. This background
has been obtained using a thin filter which is different from the thick filter of our observation.
However, it does not affect the lower energy spectrum because it is more than one order
of magnitude below Kepler's SNR spectrum at 0.3 keV.
For the following spectral studies (see Sect.\ref{partie spectrale}), we have used
the method which consists of calculating a weight for each photon to correct for the vignetting effect.
The chosen on-axis response files are \texttt{m$*$\_thickv9q20t5r6\_all\_15.rsp} for MOS and \texttt{epn\_lw20\_sdY9\_thick.rsp} for pn.

The EPIC spectra extracted from the entire remnant are shown in 
Figure \ref{spectre_all}.
These X-ray spectra are expected to be the combination of at least two 
optically thin thermal continua plus emission lines coming from 
highly ionized atoms of the ejecta. The first continuum arises from the shocked ejecta and 
the second continuum comes from the shocked ambient medium.
A nonthermal component (due to synchrotron process or nonthermal bremsstrahlung) may also be present. 
Chiefly, He $\alpha$ features of silicon, sulphur, argon, calcium and the iron K$\alpha$ line
dominate the EPIC spectra from about 1.4 keV. Below, numerous transitions 
of the iron L lines are blended with magnesium, neon and oxygen K lines.
At first sight, the overall spectrum resembles Tycho's (Decourchelle et al. 2001). However, it is noteworthy that
the Fe L complex and the Fe K line respectively dominate over the Si and the Ca lines, unlike in Tycho.

\begin{table}[t]
\centering
\begin{tabular}{lc}
\hline
Spectral feature & Energy band (keV) \\ \hline \hline
Fe L (Fe \textsc{xvii}) & 0.78 - 0.86 \\ \hline
Continuum below line & 1.42 - 1.50 \\
Si \textsc{xiii} He$\alpha$ & 1.75 - 1.94 \\
Continuum above line & 2.03 - 2.09 \\ \hline
4-6 keV continuum & 4.17 - 5.86 \\
Fe K$\alpha$ & 6.18 - 6.69 \\
8-10 keV continuum & 8.10 - 10.00 \\ \hline
\end{tabular}
\caption{Energy cuts for images.}
\label{bands}
\end{table}

The emission-line images are formed from these spectra by selecting narrow energy bands.
Table \ref{bands} gives the energy cuts chosen for these images.
The width of the Si K and Fe K energy bands have been optimized (best signal-to-noise ratio) by a method explained in Appendix \ref{appendix1.1}.
We also cleaned the images contaminated by continuum contribution with an interpolation based method which is described in Appendix \ref{appendix1.2}.
Above all, this was done for the Fe K image using the surrounding continuum, which is well represented with an exponential model.
The method was also applied to the Si K using a power-law model.
However, one should keep in mind that this method is approximate since 
the base level around the Si K line does not correspond exactly to the continuum level.
This method was not applied to the Fe L line owing to the fact that
the continuum level is not well known around this line.

The mapped lines are those that have the highest signal-to-noise ratio above the continuum:
Si K$\alpha$, Fe L and Fe K$\alpha$.
We do not show maps of fainter lines of S, Ca, and Ar that are present in the integrated spectrum.
They are roughly similar to the map of the Si K line emission.

\begin{figure*}[t]
\centering
\includegraphics[width=16cm,height=0.2cm]{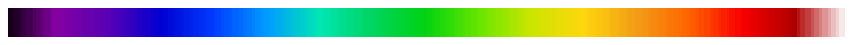}
\begin{tabular}{cc}
\includegraphics[width=8cm]{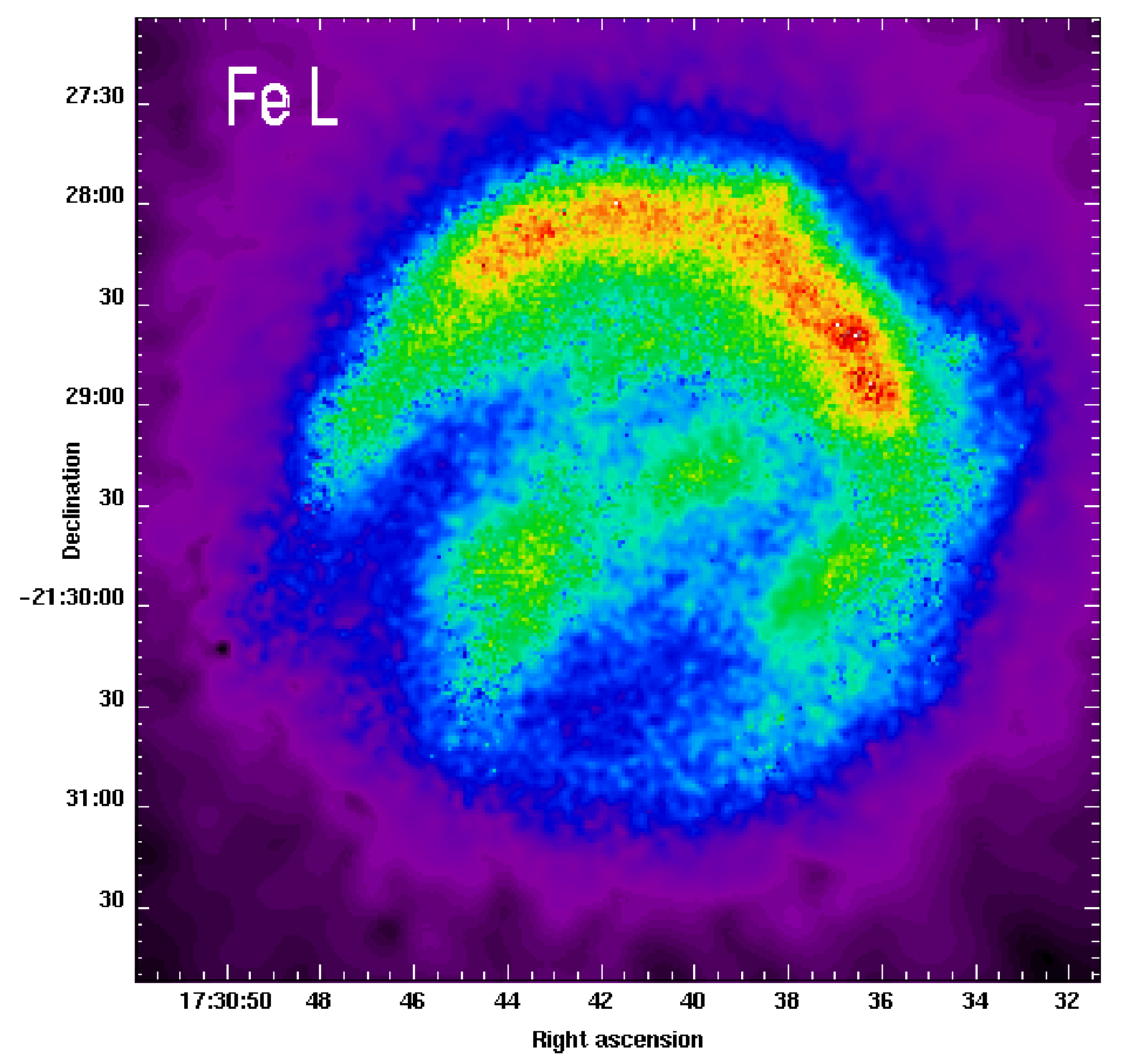} & \includegraphics[width=8cm]{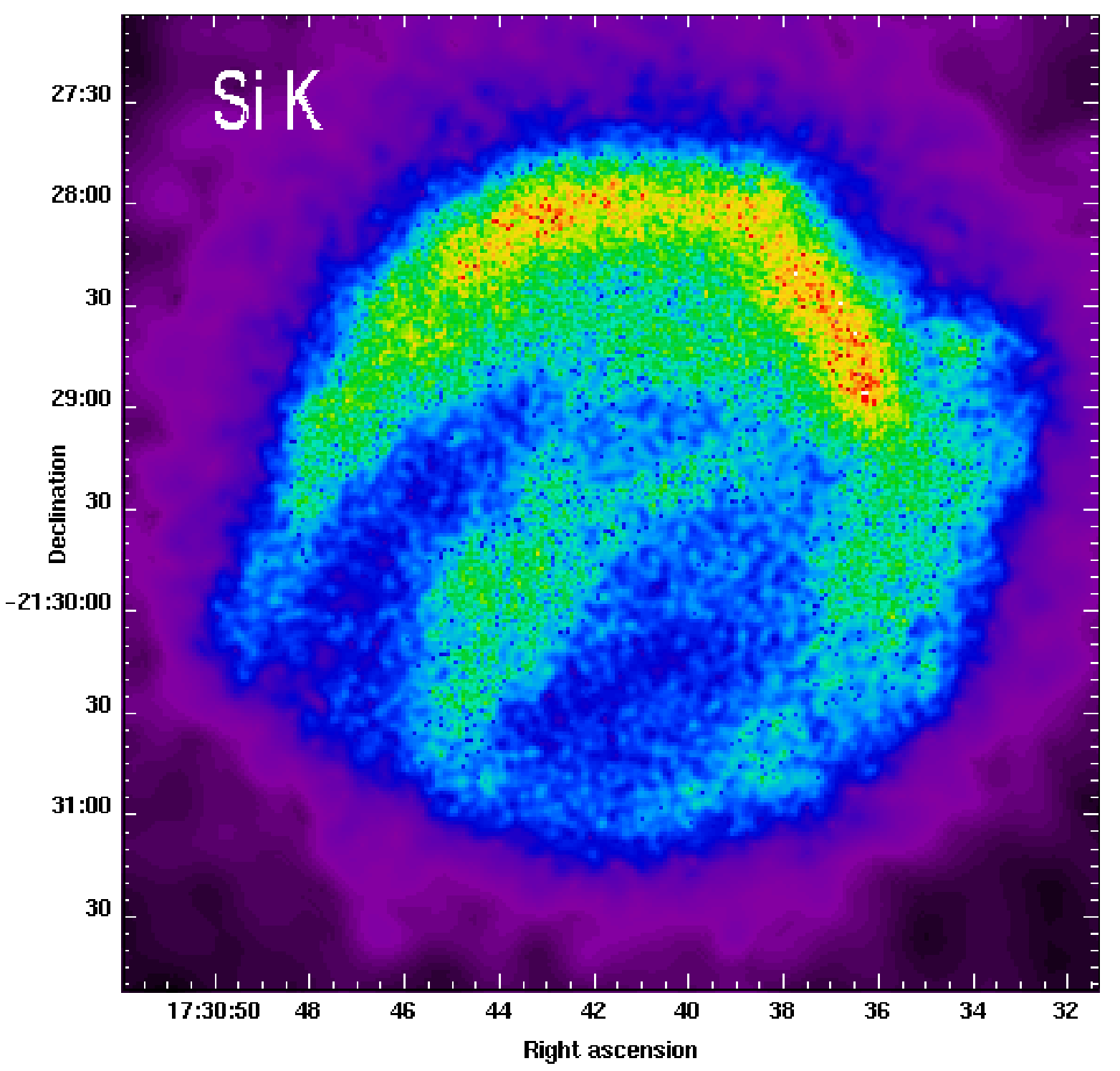} \\
\includegraphics[width=8cm]{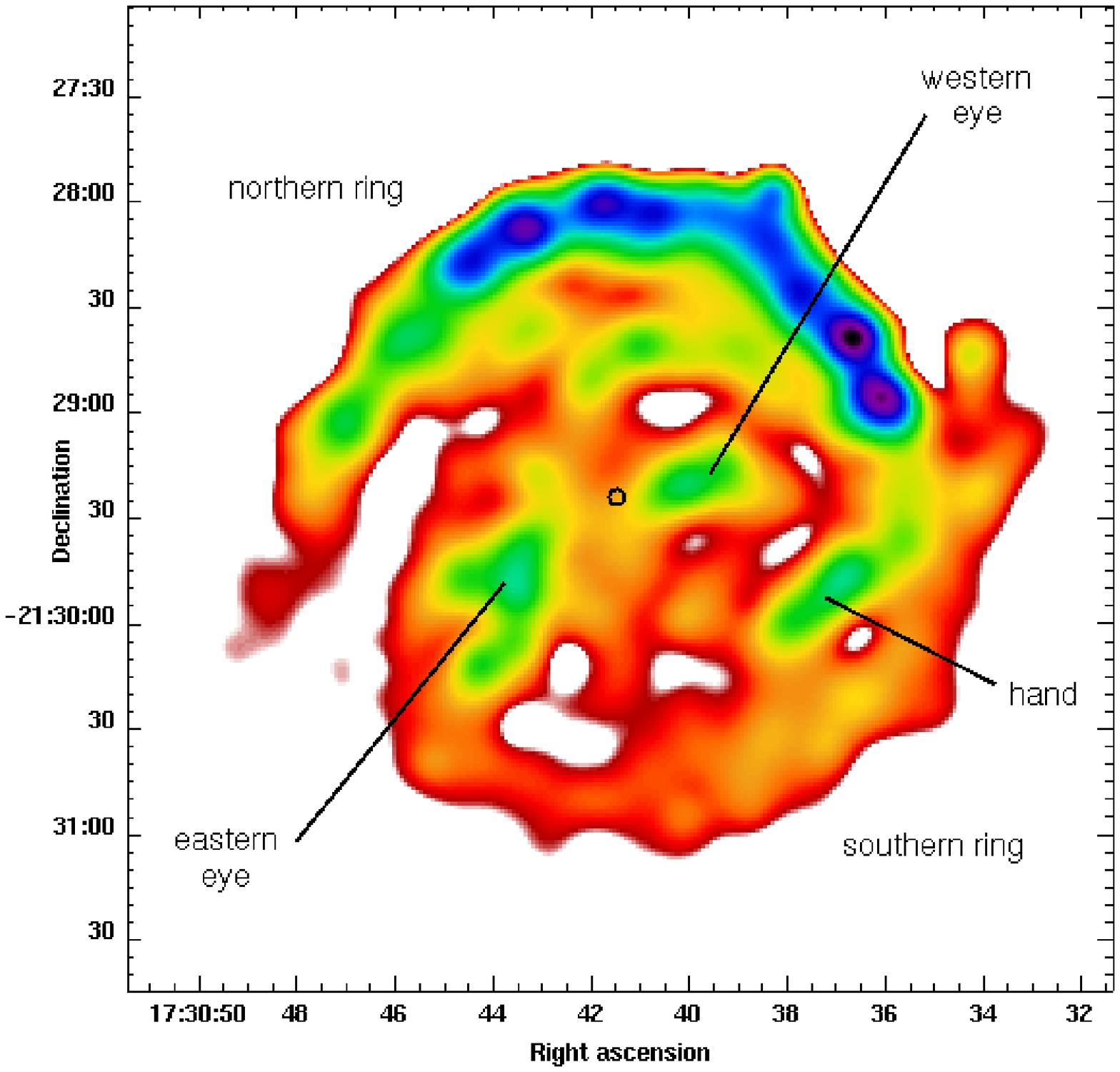} & \includegraphics[width=8cm]{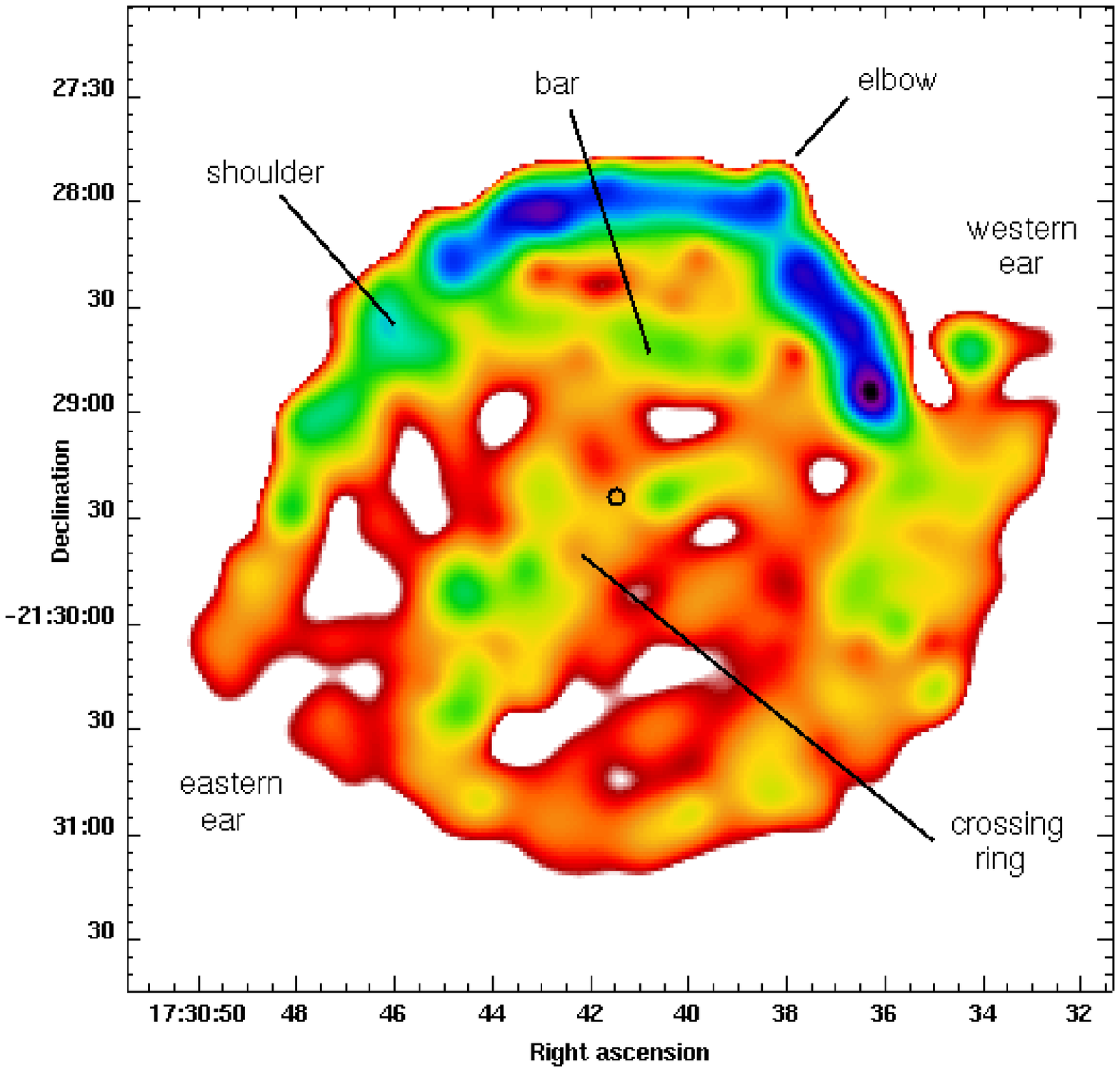}
\end{tabular}
\includegraphics[width=16cm,height=0.2cm,angle=180]{3634_f2.eps}
\caption{\textit{Top-left panel}: MOS vignetting corrected count rate image of Fe L (780-860 eV band). The data are smoothed with the same template used for the Si K image (Top-right).
\textit{Bottom-left panel}: Fe L deconvolved image in the same band.
\textit{Top-right panel}: MOS vignetting corrected count rate image of Si K (1755-1940 eV band). The data are adaptively smoothed to a signal-to-noise ratio of 5. 
\textit{Bottom-right panel}: Si K deconvolved image in the same band. All images use a square-root scaling. The small black circle in both deconvolved
images indicates the center chosen to construct the radial profiles.}
\label{image Fe Si}
\end{figure*}

\begin{figure*}[t]
\centering
\begin{tabular}{cc}
\includegraphics[width=8.5cm]{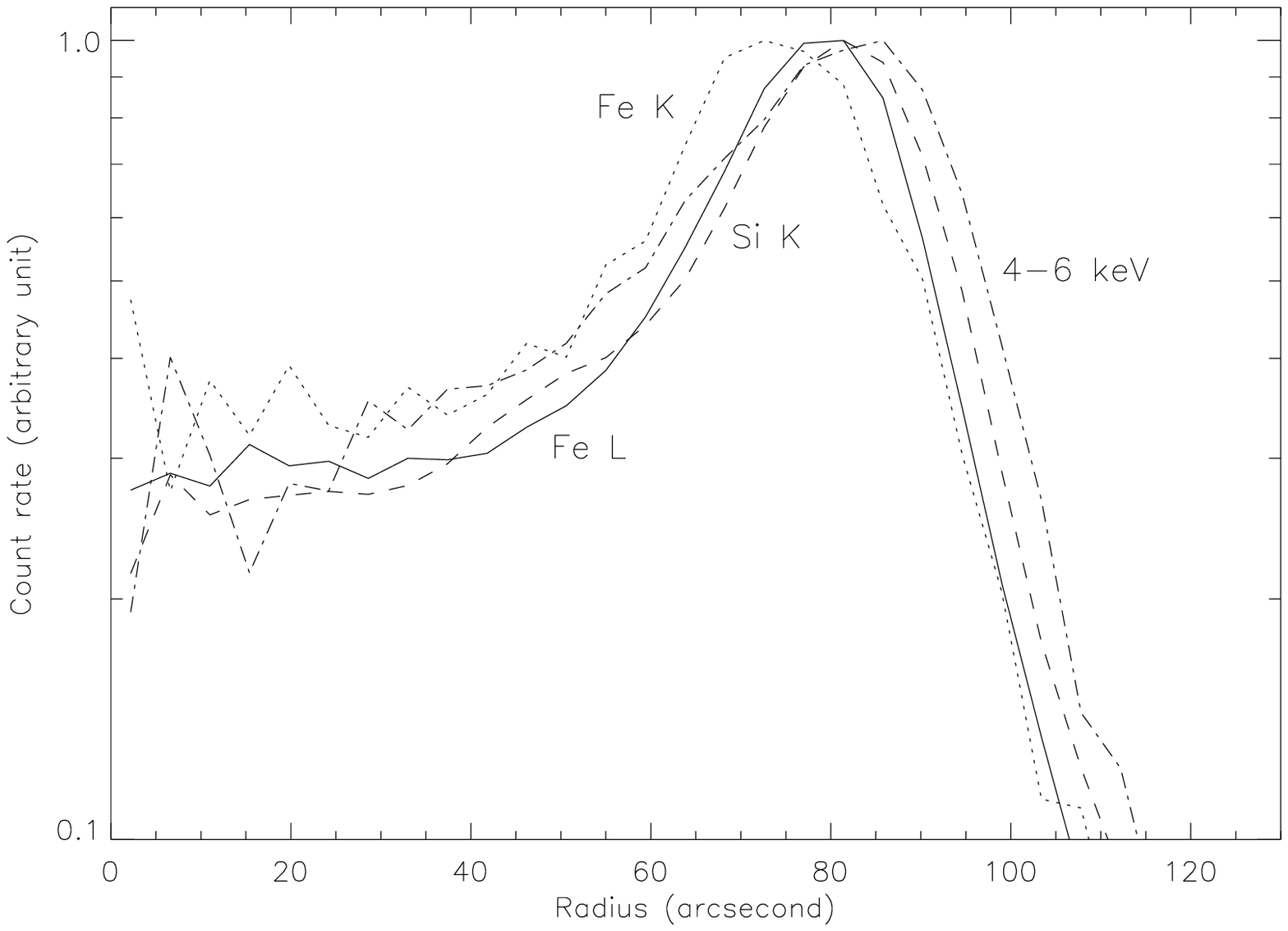} & \includegraphics[width=8.5cm]{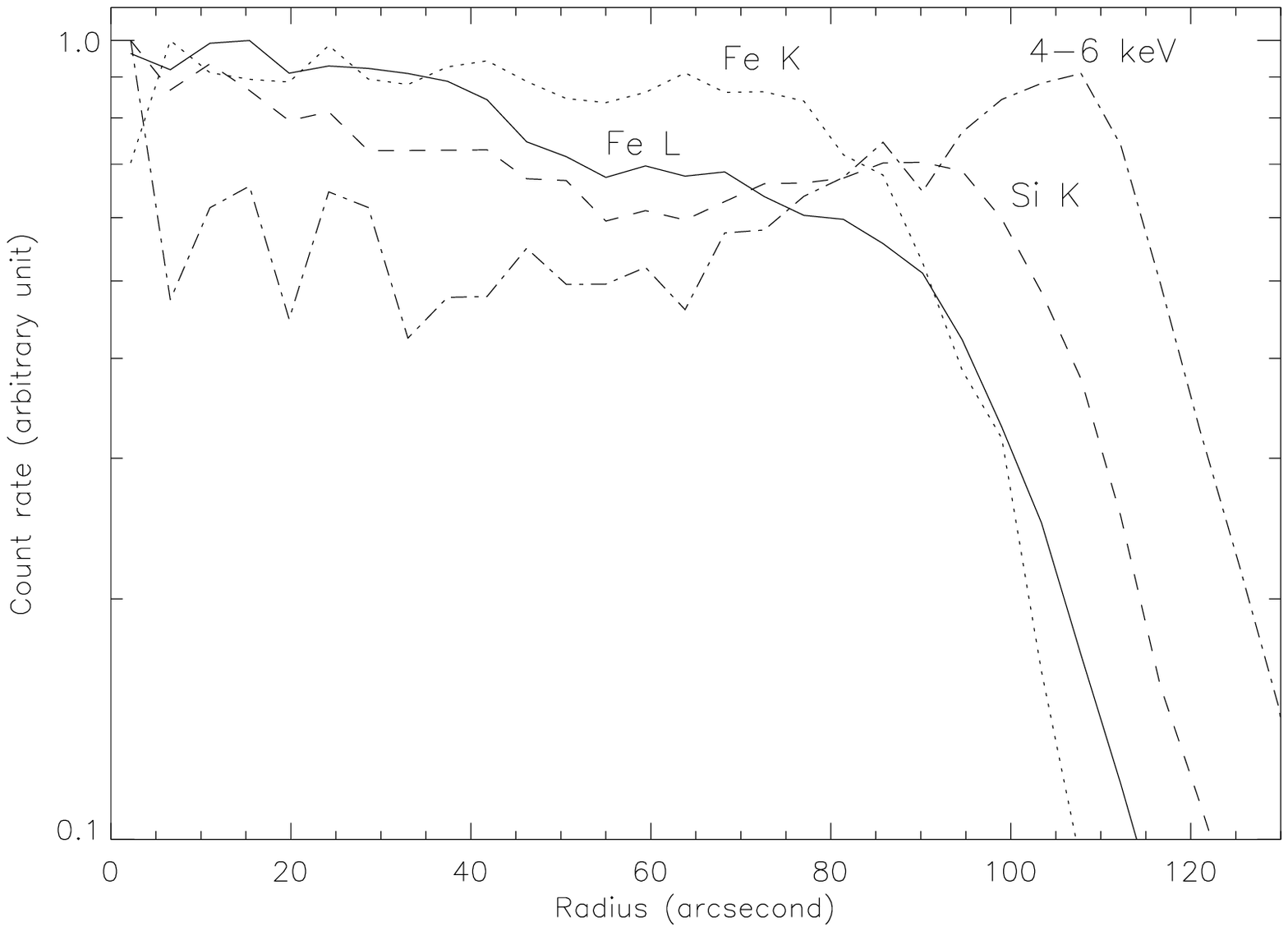}
\end{tabular}
\caption{\textit{Left panel}: MOS+pn radial profiles of 
Fe L (solid line), Si K (dashed), Fe K (dotted) and 4-6 keV continuum (dash dot) 
between $10^\circ$ and $170^\circ$ from west (northern ring). 
The coordinates of the center ($17$h$30$m$41$s,$-21^\circ 29 \arcmin 23 \arcsec$) are taken from Hughes (1999).
It is marked by the small black circle on Figure \ref{image Fe Si}.
The radial profiles were built from the vignetting corrected count rate images of Fe L, Si K and 4-6 keV continuum
(background subtracted) and Fe K (continuum and background subtracted) with a step radius of $4.4 \arcsec$.
Each radial profile has been renormalized to its maximum. 
\textit{Right panel}: same as left panel but between $175^\circ$ and $360^\circ$.
}
\label{rp}
\end{figure*}

\section{Emission-line images}
\label{elt images}

Figure \ref{image Fe Si} (top) shows Fe L and Si K adaptively smoothed images.
All adaptive smoothings were made with the asmooth package of the XMM SAS.
Figure \ref{image Fe Si} (bottom) shows their respective deconvolved images to better illustrate the main features of the SNR.
Deconvolved images have been constructed using the MOS PSF at 1500 eV (established from the calibration files) and
the method is the regularized gradient method with 40 iterations, assuming the noise to be Poisson (\cite{starck}).
The FWHM of the PSF is 4.4 arcseconds at 1500 eV.

In young SNRs, we expect to see at least two different emission zones coming on one hand from the shocked ejecta and 
on the other hand from the shocked ambient medium. 
The X-ray emission, crudely proportional to the density squared, mainly arises from the shocked ejecta which are usually denser
than the ambient medium and enriched in heavy elements.
Thus, the brightest emission lines give information on the ejected supernova matter.

The comparison between the spatial distribution of Fe L and Si K indicates an overall morphological agreement both
in the almost spherical diffuse emission and in some of the brightest regions. Indeed, the north-south brightness
asymmetry of the ring, the bar, the elbow, the western ear, and the eastern eye are shared by both images.
A closer look reveals that some differences exist notably at the eastern ear where there seems
to be more Si K than Fe L and at the hand where it is the contrary. 
We have constructed separately the radial profiles of the northern ($10^\circ$-$170^\circ$, from west anti-clockwise) 
and southern ($175^\circ$-$360^\circ$) parts of Kepler's SNR from the images.
These radial profiles (Fig. \ref{rp}) emphasize the strong contrast between the north (left panel) and the south (right panel). 
The north emission is exclusively dominated by the northern ring 
whereas at the south, a large part of the inner emission is due to the eastern eye.
We also see that both Fe L and Si K spread out further in the south than in the north.
These radial profiles show that the Si K emission peaks at a larger radius than does Fe L: slightly in the north
and rather more in the south.
It shows that there is no inversion of the Si and Fe layers as detected in Cas A (Hughes et al. 2000; Hwang et al. 2000; Willingale et al. 2002).
These clues agree with an onion layer structure (at least for iron and silicon) which would have persisted after the explosion. 

\begin{figure}[t]
\centering
\includegraphics[width=8cm]{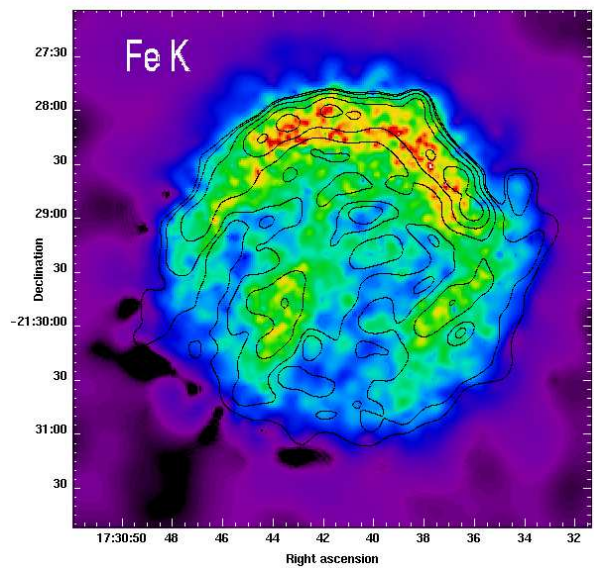}
\caption{Fe L contours of the deconvolved image superimposed with the MOS+pn Fe K vignetting corrected count rate image (continuum and background subtracted)
in the 6180-6169 eV band. The data are adaptively smoothed to a signal-to-noise ratio of 5. The intensity scaling is square-root.}
\label{image ferK}
\end{figure}

The comparison between the Fe L and Fe K emission can give information on the
temperature structure inside the remnant and on the ionization timescale. 
Figures \ref{rp} and \ref{image ferK} illustrate that the Fe K emission peaks at a smaller radius than the Fe L 
emission in the north as well as in the south.
This tells us that the inner ejecta should be hotter than the outer ejecta, but not directly by how much.
Kinugasa \& Tsunemi (1999) have modelled the integrated spectrum of Kepler's SNR with the \textit{ASCA} data
using two non-equilibrium thermal components with the same abundances for both components but with different temperatures.
They found a temperature of $0.36 \pm 0.04$ keV and $2.26 \pm 0.15$ keV for the Fe L and Fe K, respectively.
Even though the Fe K was underproduced with this two-component model,
this provides evidence for a strong temperature gradient (a lower limit temperature ratio of $\sim 6$) in the shocked ejecta medium turned inwards of the remnant.
We note that an ionization effect would work in the opposite direction. The ratio between the Fe K and the Fe L line strengths
should increase when going away from the reverse shock as Fe gets more ionized.

The temperature increase inwards must be the result of the remnant's hydrodynamics.
Hydrodynamical models give the profiles of pressure $P$ and mass density $\varrho$ as a function of the radius.
We note that the profile of $P / \varrho$ is equivalent to the temperature profile if the mean molecular weight remains constant, \emph{i.e.}
if the abundance of the dominant elements does not vary in the remnant.
In the following paragraphs, we assume constant elemental composition 
but we should keep in mind that any variation of the ejecta composition of the dominant species should alter our interpretation.

Models of SNRs evolution depend on the initial outer density profiles
in the ejecta and the ambient medium. 
We first assume that the supernova exploded into a constant density medium.
Two simple cases have been under consideration in the framework of spherical models.
On one hand, the ejecta density profile may follow a power-law distribution with a constant core
(\cite{chevalier2}) which is appropriate to a type II classification (\cite{arnett}). 
In this hypothesis, contrary to the observations, the ejecta are hotter near the contact discontinuity 
(interface between the shocked ejecta and the shocked ambient medium) than at the reverse shock. 
On the other hand, an exponential density profile aims to represent a type Ia progenitor (\cite{dwarkadas}). 
In this latter hypothesis, the shocked ejecta are gradually colder towards the contact discontinuity as is observed,
but the temperature profile is much too flat to account for the required two temperatures of the Fe K and Fe L ejecta (temperature ratio of $\sim 6$).
The temperature gradient of the reverse shocked region can be larger when the exponential ejecta density structure is running into a circumstellar medium (\cite{dwarkadas})
and is then compatible with the observations\footnote{We note that a power-law ejecta density structure running 
into a circumstellar medium would also lead to a temperature slope compatible with the observations but
with a too low temperature to account for the Fe K line (Decourchelle \& Ballet 1994). So this case is rejected and no more considered in the following.}.

In the remnant of Tycho's type Ia supernova, the Fe K emission peaks at a smaller radius than the Fe L (Decourchelle et al. 2001; Hwang \& Gotthelf 1997)
as is observed in Kepler's SNR.
Both remnants share the property of a higher temperature toward the center in the shocked ejecta medium with a temperature ratio of at least $2$ 
(Hwang et al. 1998; Kinugasa \& Tsunemi 1999).
In the framework of an exponential ejecta profile, this requires an interaction with a circumstellar medium.
There is no observational indication for such a circumstellar medium around type Ia supernovae but as discussed in Dwarkadas \& Chevalier (1998),
it has been proposed that in some binary models of type Ia the mass loss of the companion can modify the environment of the progenitor.

In Kepler's SNR, there is evidence for the presence of a circumstellar material
which has been interpreted in terms of a massive progenitor (Bandiera 1987; Borkowski et al. 1992).
If the supernova was a type Ib as suggested by Borkowski et al. (1992) and has a power-law outer density profile, 
the observed higher temperature inwards the remnant would require that the reverse shock has already entered the central plateau.
This condition requires that the ejecta mass $M_{\mathrm{ej}}$ is lower than $3 M_{\odot}$ taking the parameters\footnote{$E_{\mathrm{SN}}(M_{\odot}/M_{\mathrm{ej}})^{2/3} = 0.53 \; 10^{51}$ ergs and $\rho_0 = 0.42$ amu cm$^{-3}$ 
for a power-law index of the initial outer density profile in the ejecta $n=9$ and a flat ambient medium.} derived from the best fit to
the \textit{ASCA}+\textit{XTE} data (Decourchelle et al. 2000).
If $M_{\mathrm{ej}} > 3 M_{\odot}$, the reverse shock would still be in the power-law envelope, and would give the wrong behavior in Fe K/Fe L.

The condition that the reverse shock has entered in the plateau sets an upper limit on the expansion rate of the SNR.
Indeed from the self-similar model of Chevalier (1982), the expansion parameter is $m = (n-3)/n \propto \log R / \log t$ 
assuming a power-law density profile of index $n$ expanding into a flat ambient medium.
At most $n$=12 imposes the upper limit $m = 0.75$ for a flat ambient medium.
This value is contradictory with the results of Hughes (1999) who measures a dynamical evolution near free expansion ($m \sim 0.93$).
However, it is consistent with the highest value $m \sim 0.65$ found from the radio measurement (Dickel et al. 1988).

\section{High energy continuum}
\label{high energy continuum}

\begin{figure}[t]
\centering
\includegraphics[width=8cm]{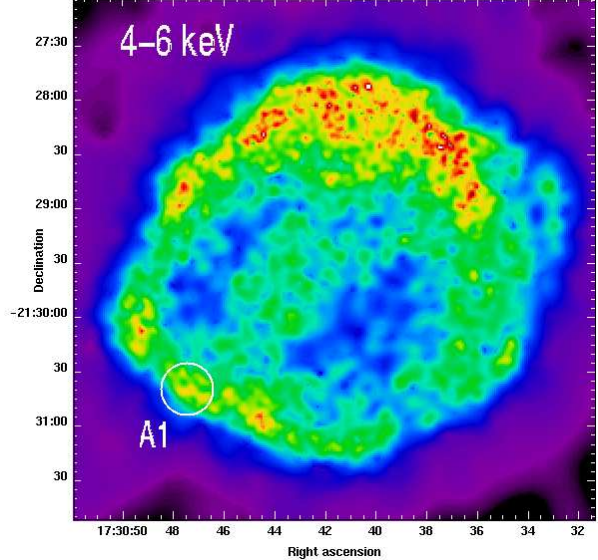}
\caption{MOS+pn vignetting corrected count rate image of the 4-6 keV continuum. 
The data are adaptively smoothed to a signal-to-noise ratio of 5.
The intensity scaling is square-root. 
The spectrum of the region A1 is shown in Figure \ref{synchrotron}.
The center of the region A1 is ($17$h$30$m$47$s,$-21^\circ 30 \arcmin 39 \arcsec$), its
radius is $\sim 14 \arcsec$ and it is $\sim 105\arcsec$ away from the SNR center.}
\label{c46_a10}
\end{figure}

\begin{figure}[t]
\centering
\begin{tabular}{cc}
\includegraphics[width=4cm]{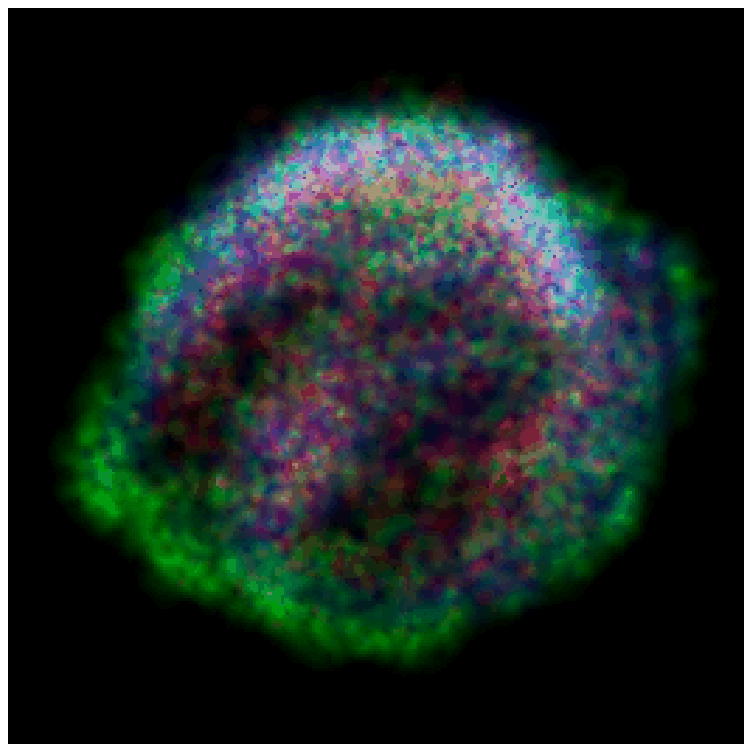} & \includegraphics[width=4cm]{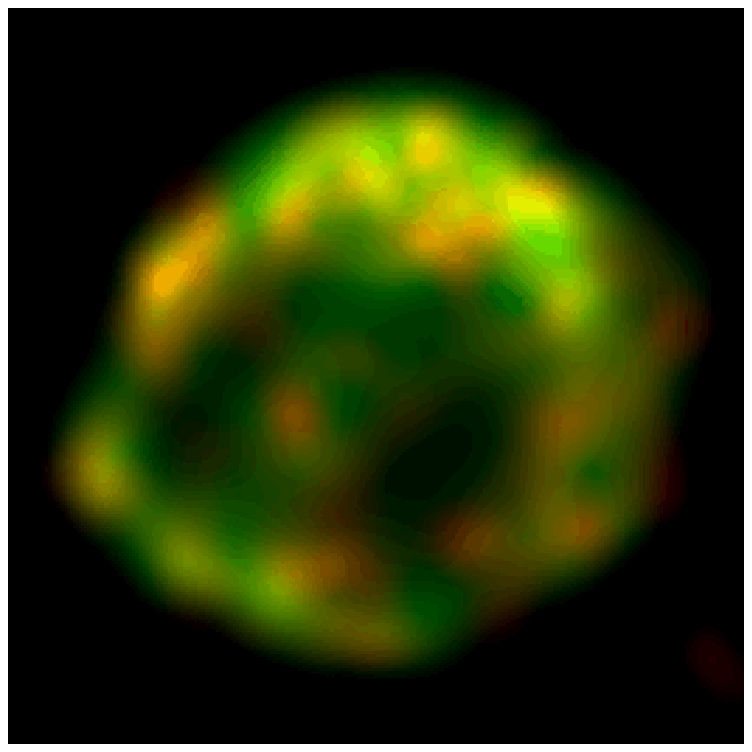}
\end{tabular}
\caption{\textit{Left panel}: Three color image of the Si (blue), 
Fe K (red), 4-6 keV high energy continuum (green). 
All images are vignetting corrected and are smoothed with the template 
created from the adaptively smoothed count image of the \mbox{Fe K} + \mbox{4-6} keV (signal-to-noise of 5).
\textit{Right panel}: 
MOS+pn two color image of the high energy continuum:
4-6 keV map in green, 8-10 keV map in red,
common areas are in yellow. Both count rate images are background subtracted, vignetting corrected and are smoothed with the template
created from the adaptively smoothed 8-10 keV band count image (signal-to-noise of 5). Maps use a square-root scaling
with minimum set to 30\% of maximum.\label{c46810}}
\end{figure}

As we see in Figure \ref{c46_a10}, the 4-6 keV continuum emission globally looks like the emission-line images except
in the southeast where there is a clear limb-brightened profile which broadly resembles the 21 cm radio image (Dickel et al. 1988; DeLaney et al. 2002).
Radial profiles (Fig. \ref{rp}) and the three color image (Fig. \ref{c46810} left panel)
indicate that the continuum emission occurs at a larger radius than all the line emission, especially in the southeast.
We surmise that the outermost continuum emission is the signature of the blast wave.
The north emission is the strongest. 
The southeastern side (near the eastern ear) defines a ridge of emission dominated 
by the presence of a few bright features at \textit{XMM-Newton} spatial resolution. 
These features seen with \textit{XMM-Newton} correspond to thin curved shocks observed in the \textit{Chandra} image.
The spectrum of the region A1 (cf. small circle in Fig. \ref{c46_a10}) will be analysed in Sect. \ref{partie spectrale}.
A diffuse emission region exists at the center of Kepler's SNR toward the eastern eye. 
The southwestern edge is less pronounced than the other outer rims.

Figure \ref{c46810} (right panel) shows that the 4-6 keV and 8-10 keV bands correspond well and that the continuum emission essentially arises from
an outer ring. 
It also illustrates 
how the hardness of the continuum changes all along the remnant
and gives locally an indication of the shape of the high energy spectrum. 
Regions where the continuum is flat (in orange) are found in the
north, at the shoulder and in the southeast.
It is possible that these regions are responsible for the higher energy emission (up to 20 keV) 
discovered by RXTE (\cite{decour2}).

\begin{figure*}[t]
\centering
\begin{tabular}{cc}
\includegraphics[width=8.5cm]{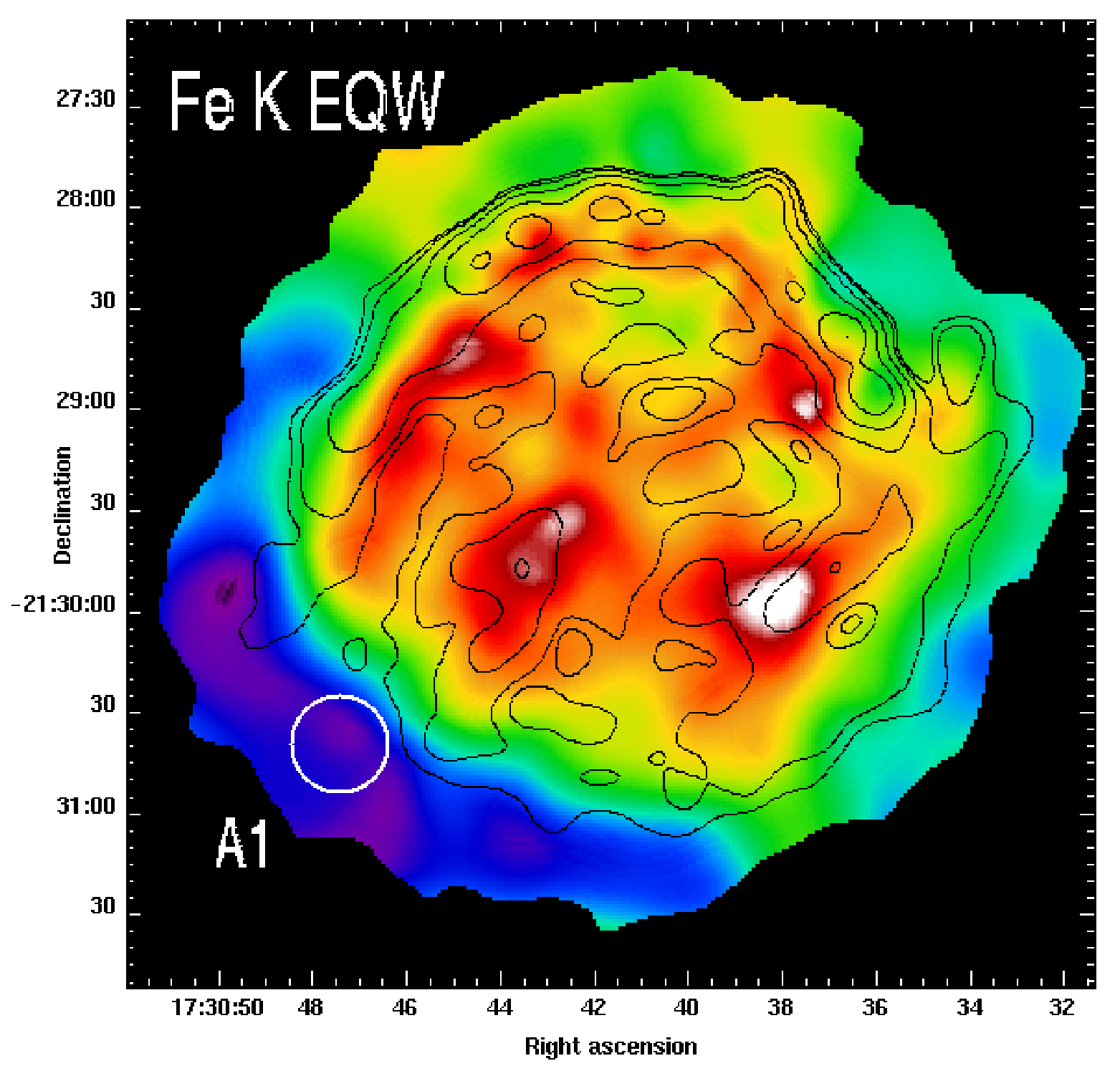} & \includegraphics[width=8.5cm]{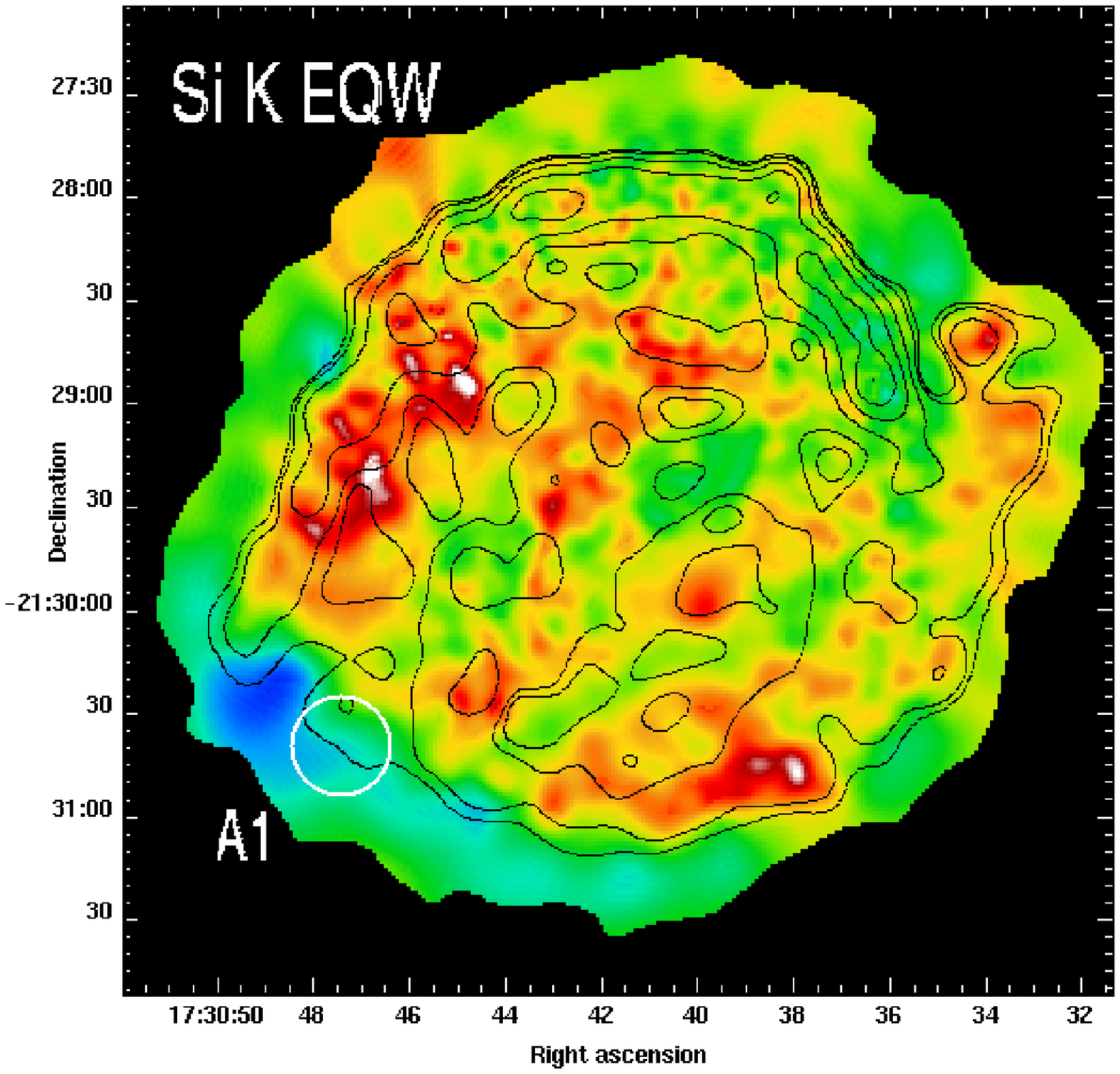}
\end{tabular}
\includegraphics[width=16cm,height=0.2cm]{3634_f2.eps}
\caption{\textit{Left panel}: MOS+pn equivalent width image of the Fe K emission. 
This image uses a square-root scaling. Colorbar is in keV: purple-blue=0-1, green=1-2.7, yellow-orange=2.7-4.5, red=4.5-6.7, white=6.7-7.4.
The ratio refers to images smoothed with the template created from the adaptively 
smoothed count image of the \mbox{Fe K} + \mbox{4-6} keV (signal-to-noise of 19).
The superimposed contours (sqrt scale) come from the MOS Fe L deconvolved image. 
\textit{Right panel}: MOS equivalent width image of the Si K emission. 
This image uses a square-root scaling. Colorbar is in keV: blue=0-0.2, green=0.2-0.7, yellow-orange=0.7-1.1, red=1.1-1.7, white=1.7-2.0
The ratio refers to images smoothed with the template created from the adaptively 
smoothed count image of the \mbox{Si K} + \mbox{1.42-1.50} keV (signal-to-noise of 19).
The superimposed contours (sqrt scale) come from the MOS Si K deconvolved image.
Both images indicate the location of the bright region A1 seen in the 4-6 keV high energy continuum image (see Fig. \ref{c46_a10}).}
\label{eqw}
\end{figure*}

The northern part of Kepler's SNR is the brightest in the silicon and iron maps as well as in the
high energy continuum map (see for example Fig. \ref{c46810}, left panel).
At first sight, it may be due to a density enhancement in the ambient medium toward the north. 
Since brightness is proportional to the square of the density,
such an enhancement would explain the strong brightness of the northern rim in all 
the different energy bands that we have selected to construct our images.
The southeastern emission (from the shoulder to the southern ring), which is faint in the emission lines,  
may essentially be due to a low ambient medium density. 
To check these assumptions, it is possible to map the ratio
of the line emission to the underlying continuum, \emph{i.e.} to construct equivalent width maps.

\section{Equivalent width images}\label{EQW images}

We focus on the Fe K and Si K equivalent width (hereafter EQW) for which it
is possible to define an adjacent continuum.
The EQW map reveals the ejecta abundances but depends also on the temperature and 
on the ionization age.
Note that it also assumes that the underlying continuum is related to the lines.
To construct an EQW image, we have divided the image in the line
(background and continuum subtracted) by the 
underlying continuum (background subtracted) [see Appendix \ref{appendix1.3}].

Figure \ref{eqw} (left panel) shows the Fe K EQW.
What we see is the contrast between the center of the remnant and the exterior. The former is
associated with the ejecta (high EQW) and the latter with the forward shock (weak EQW). 
Hence, this confirms that the iron originates from the ejecta. 
We note that the Fe K ejecta are mainly found at the shoulder, the eastern eye and below the hand (high EQW).
On the other hand, the southeastern region is extremely faint in Fe K (weak EQW) but bright in the high energy continuum.

When focusing our attention in the inner part of the ejecta (in red, Fig. \ref{eqw}, left panel), we see that
there is no obvious correlation between the EQW map and the flux in the Fe K ejecta. 
In particular the northern rim is no longer dominant in brightness. This is an indication that
abundance variations are not the cause of the north-south brightness asymmetry
and implies some differences in the density of the ambient medium around Kepler's SNR.
This point is consistent with the presence of dense optical knots in the northern ring (\cite{bandiera2}).

Figure \ref{eqw} (right panel) shows the Si K EQW.
The contrast between the center of the remnant and the exterior is less pronounced than in the Fe K EQW.
The Si K ejecta are found towards the shoulder, near the southwestern edge, at the crossing ring and at the western ear (in yellow and red).

Firstly, it shows afresh that the Si K ejecta have been dispersed to a higher radius than the Fe K ejecta, notably in the south.
Secondly, the northern rim is not strong and confirms again that the north-south brightness asymmetry is not caused by abundance variations.
As for the southeastern side of the remnant, it presents a lower contrast than in the Fe K EQW 
because the Si K extends to a larger radius than the Fe K as shown in the radial profiles (Fig. \ref{rp}, right panel).
While the Si K ejecta emission is present in the western ear, the eastern ear does not show evidence for the presence of ejecta. 

The search for prominent interstellar H\textsc{i} features around Kepler's SNR has lead to negative result (Reynoso \& Goss 1999).
Thus, it is likely that the presence of overdensities in the north is circumstellar in origin. 
This is supported by the detection of slow-moving optical knots ($\sim 200$ km s$^{-1}$) there that have
nitrogen overabundances higher than the solar value by a factor 4 (Dennefeld 1982; Blair et al. 1991).

\section{Spectral study of the forward shock}
\label{partie spectrale}

\begin{figure*}[t]
\centering
\begin{tabular}{cc}
\includegraphics[width=8.5cm,angle=0]{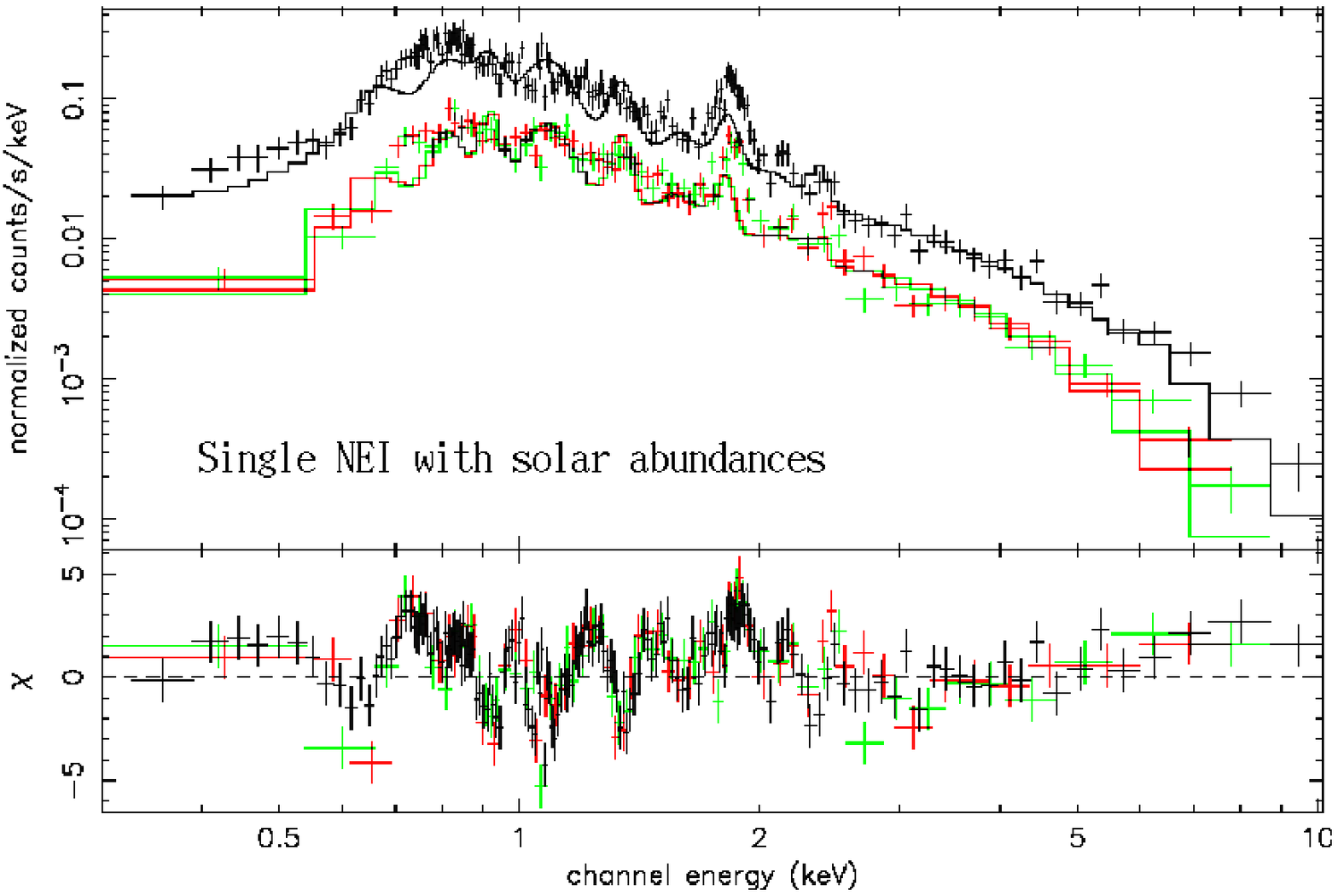} & \includegraphics[width=8.5cm,angle=0]{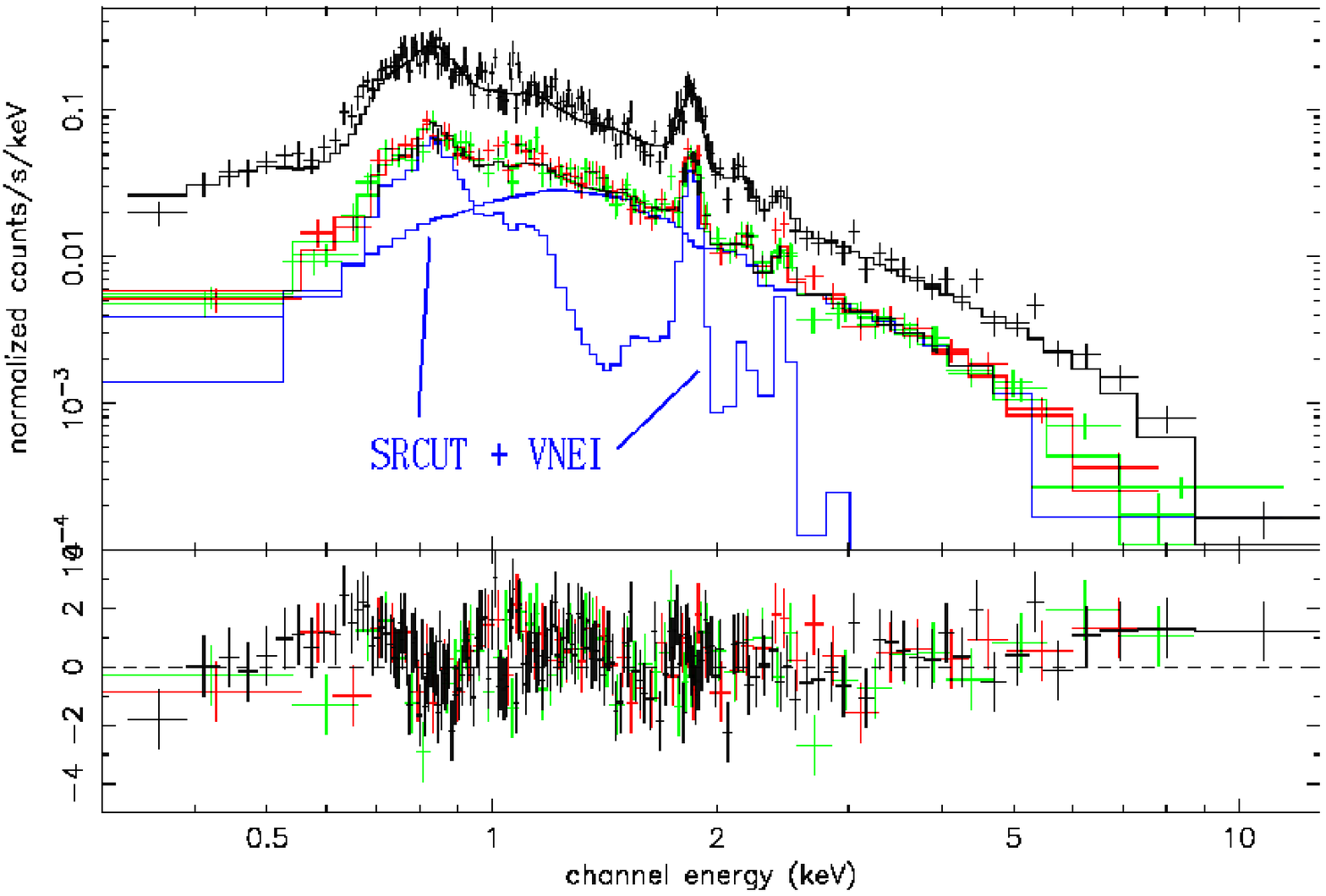}
\end{tabular}
\caption{EPIC spectra of the southeastern part of Kepler's SNR (MOS1 in green, MOS2 in red and pn in black).
\textit{Left panel}: Best-fit obtained with the single NEI model
where all the elemental abundances are fixed to solar abundances (see Sect. \ref{1 nei} and Table \ref{res_XSPEC_1}, first model).
\textit{Right panel}: Best-fit obtained with the SRCUT+NEI model (see Sect. \ref{srcut+nei} and Table \ref{res_XSPEC_2}, second model).
For this model, we have illustrated the decomposition of the different components for MOS1 only. 
Both images shows the EPIC spectra rebinned to 5$\sigma$ to better distinguish the different components
but all the spectral analysis was made with a threshold at 3$\sigma$.}
\label{synchrotron}
\end{figure*}

In this section, we intend to characterize the spectrum of the southern regions dominated by strong
high energy continuum (4-6 keV and 8-10 keV) and weak EQW of Si K and Fe K.
For these regions, we attribute the X-ray emission to the shocked ambient medium located just behind the blast wave.
We will not focus in detail on the modelling of the contribution of the shocked ejecta which arises in our spectrum. 
Note that a detailed study of the forward shock all along the southeast rim would require \textit{Chandra} spatial resolution.

Here, we will present the first attempt to characterize the spectrum associated with the forward shock in Kepler's SNR.
For this purpose, we have extracted the EPIC spectra (Fig. \ref{synchrotron}) 
of the bright region A1 in the southeastern area (see Fig. \ref{c46_a10}).
These spectra have a continuum extending till 10 keV with little line emission (Fig. \ref{synchrotron})\footnote{We grouped the channels of these spectra until we have a signal-to-noise ratio greater than $3 \sigma$ in each bin.}.
Though spatial redistribution of emission from the brightest regions in the north could
contaminate the region A1, we find that regions one arcminute from the center of the region A1
contribute only $\sim 5\%$ of the flux in A1 at 1.5 keV.
Moreover, the spectra of other regions along the southeastern rim are qualitatively similar to A1's.

\subsection{Single non-equilibrium model}
\label{1 nei}

\begin{table}
\centering
\begin{tabular}{lll}
\hline
\multicolumn{1}{c}{Parameter} & \multicolumn{1}{c}{Solar abundances} & \multicolumn{1}{c}{Free abundances} \\ \hline \hline
$k T$ (keV) & $2.8$ (2.4-2.8) & 3.0 (2.8-3.2) \\
$\tau$ ($10^{9}$ s cm$^{-3}$) & 5.0 (4.9-5.3) & 8.6 (7.9-9.4)\\
Si (Ab$_\odot$) & $[1]$ & 1.6 (1.3-1.8) \\
S (Ab$_\odot$)  & $[1]$ & 0.4 (0.1-0.7) \\
Fe (Ab$_\odot$) & $[1]$ & 0.14 (0.12-0.16) \\
$Norm$ & $6.0 \: 10^{-4}$ (5.6-6.5) & $ 5.0 \: 10^{-4}$ (4.6-5.5)\\ \hline
$N_{\mathrm{H}}$ ($10^{22}$ cm$^{-2}$) & 0.86 (0.84-0.87) & 0.30 (0.27-0.33) \\ \hline
$\chi^2$ & 1287/607 dof & 693/596 dof \\ \hline \hline
\end{tabular}
\caption{Best-fit parameter values for the different single non-equilibrium models using the MOS and pn data all at once. 
The first model refers to the NEI model with all elements fixed to solar abundances.
The second model refers to the NEI model with all the elemental abundances (C, N, O, Ne, Mg, Si, S, Ca, Fe, Ni) freely fitted excepted H and He 
which were frozen to solar abundances.
The second model leads to abundances of C, N, O, Ne, Ca near zero.
The errors are in the range 
$\Delta \chi^2 < 2.7$ (90\% confidence level) on one parameter.
Quantities in square brackets are held fixed at the given value.}
\label{res_XSPEC_1}
\end{table}

Firstly, we have supposed that the X-ray emission of the region A1 came entirely from the shocked ambient medium.
We applied a single non-equilibrium (NEI) model to the region A1 within XSPEC (version 11.1, Arnaud 1996)
freezing all the elemental abundances to solar abundances (given by Anders \& Grevesse 1989).
In addition to the elemental abundances, this NEI model embraces the following parameters:
the temperature $k T$ and the ionization age $\tau$. We define
$\tau$ as $\int_{t_0}^{t_s} n_e \: dt$, where $n_e$ is the electron density and where
$t_s$ and $t_0$ are respectively the age of the SNR and the time when the gas was shock-heated by the blast wave
(setting zero to the explosion date).

With all the elemental abundances held fixed at solar abundances,
absorption along the line-of-sight, temperature and ionization age were freely fitted.
This results in a poor fit ($\chi^2$= 1287 for 607 dof) which overproduces the lines from
intermediate-mass elements (O, Ne, Mg) and underproduces the Si line as shown in Figure \ref{synchrotron} (left panel).
This shows that at least the Si comes from the ejecta.
Figure \ref{synchrotron} (left panel) shows also that the centroid energy position of the Si and S lines differs when comparing the model and the data.
The best-fit parameters are given in Table \ref{res_XSPEC_1} (Solar abundances model).

Then, we thawed all the elemental abundances 
(C, N, O, Ne, Mg, Si, S, Ca, Fe, Ni) except H and He 
which were frozen to solar abundances.
The fit (not shown) is much improved ($\chi^2$= 693 for 596 dof) provided that the abundances
of the elements C, N, O, Ne, Ca are close to zero.
The best-fit parameters are given in Table \ref{res_XSPEC_1} (Free abundances model).
The Si line, better represented, requires a slightly larger value than solar abundance while the S and the Fe
are required to be subsolar.
It seems that the Fe is necessary for representing some portion of the low energy spectrum,
although we do not have the spectral resolution there with the EPIC instruments to clearly identify individual Fe L lines. 

Even though the values of the temperature and the ionization age are reasonable, such a model is not
consistent with the composition of the ambient medium.
Nevertheless, it shows that the spectrum of the region A1 can be reproduced formally only with a continuum and lines of Si, S and Fe. 
So, we replaced the continuum arising from the H and He by a second component.

\subsection{Two components models}

\begin{table}
\centering
\begin{tabular}{lll}
\hline
\multicolumn{1}{c}{Parameter} & \multicolumn{1}{c}{Thermal model} & \multicolumn{1}{c}{Nonthermal model} \\ \hline \hline
$k T_1$ (keV) & $2.7$ (2.5-2.9) & \multicolumn{1}{c}{-} \\
$\tau_1$ ($10^{8}$ s cm$^{-3}$)  & $1.0$ ($<1.3$) & \multicolumn{1}{c}{-} \\
$\alpha$ & \multicolumn{1}{c}{-} &  0.66 (0.65-0.67) \\
$\nu_r$ ($10^{17}$ Hz) & \multicolumn{1}{c}{-} & 2.1 ($1.7$-$3.0$) \\
$Norm_1$ & $5.6\: 10^{-4}$ ($5.2$-$6.0$) & $[0.16 \: \mathrm{Jy}]$  \\ \hline
$k T_2$ (keV) & $0.56$ (0.54-0.58) & $0.54$ (0.52-0.57) \\
$\tau_2$ ($10^{11}$ s cm$^{-3}$) &  $3.3$ ($2.0$-$10.$) & $3.3$ ($>1.8$) \\
Si (Ab$_\odot$) & $[1]$ & $[1]$ \\
S (Ab$_\odot$) &  0.8 (0.4-1.2) & 1.3 (0.9-1.7) \\
Fe (Ab$_\odot$) & 0.10 (0.09-0.12) & 0.11 (0.09-0.13) \\
$Norm_2$ & $1.5 \: 10^{-3}$ ($1.3$-$1.8$) & $1.8 \: 10^{-3}$ ($1.5$-$2.0$)\\ \hline
$N_{\mathrm{H}}$ ($10^{22}$ cm$^{-2}$) & 0.28 (0.25-0.30) & 0.35 (0.33-0.37) \\ \hline
$\chi^2$ & 701/601 dof & 627/602 dof \\ \hline \hline
\end{tabular}
\caption{Best-fit parameter values for the different two components models using the MOS and pn data all at once.
The subscript 1 refers to the hot component (\emph{i.e.} shocked ambient medium) 
and the subscript 2 to the cold component (\emph{i.e.} shocked ejecta). The errors are in the range 
$\Delta \chi^2 < 2.7$ (90\% confidence level) on one parameter.
Quantities in square brackets are held fixed at the given value.}
\label{res_XSPEC_2}
\end{table}

We introduce a second component to allow for the different conditions for the lines and for the continuum.
On one hand, the lines are attributed to the ejecta and thus we introduce a plasma composed only of Si, S and Fe at low temperature.
This constitutes the cold component for which we fixed the Si abundance to solar abundance and we only thawed the S and Fe elements 
while fixing all the others including H and He to null solar abundances for the cold component
(henceforth, what counts is only the abundance ratio with regard to Si).
The temperature $kT_2$ and the ionization age $\tau_2$ were freely fitted.
On the other hand, the continuum is attributed to the shocked ambient medium and is associated to a hot temperature plasma.
For this hot component, we consider two possibilities: another NEI model or an additive nonthermal component.

\subsubsection{Thermal model}
\label{2 nei}

In this model, the shocked ejecta are modelled by a first NEI at low temperature and 
the shocked ambient medium by a second NEI at high temperature.
All the elemental abundances were fixed to solar values.
The temperature $kT_1$ and the ionization age $\tau_1$ were freely fitted.

The best-fit parameters are given in Table \ref{res_XSPEC_2} and show that the temperature is high ($\simeq 2.7$ keV)
and the ionization age very low ($\simeq 10^8$ s cm$^{-3}$) for the shocked ambient medium. 
This is very similar to what was found in Tycho's SNR (Hwang et al. 2002).
Indeed, in Tycho's SNR the temperatures of the shocked ambient medium (corresponding to \textit{Behind Rim 1,2,3,4}
in Hwang et al. 2002) are around 2 keV and the ionization age has
an upper limit in the range of a few $10^8$ s cm$^{-3}$.

The parameters derived from our spectral fit give constraints on the density via the emission measure and 
the ionization age.
In the following, we check the consistency between these two estimates.

A mean ionization age $\hat{\tau}$ can be calculated from simple hypotheses made on our extraction region (region A1).
For that region, we assume that the integrated volume along the line-of-sight is a cylinder completely filled with
shocked ejecta and shocked ambient medium.
We define the mean ionization age $\hat{\tau} = \xi \: \hat{n}_e \: t_s$ where 
$\xi$ is a parameter depending on the spatial distribution,
$\hat{n}_e$ is the mean electron density which depends on the filling factor $\varepsilon$, \emph{i.e.}
the volume fraction occupied by each medium (shocked ejecta and shocked ambient medium) in the region A1, and 
$t_s$ is the age of the SNR.
Hence, $\hat{\tau}$ is a function of $\varepsilon$ and $\xi$.
In Appendix \ref{appendix2}, we described in detail how to compute $\hat{\tau}$ and we give general formulae that can be applied to any SNR.
Here, we just give a numerical application of the formula (\ref{tau_prime}) for the region A1 in Kepler's SNR.

For the shocked ambient medium, we derive the following mean ionization age:~
\begin{equation}\label{tau2_prime}
\hat{\tau}_1 = 3.7^{+0.9}_{-0.6}\: 10^{10} \; \xi \; \varepsilon_1^{-1/2} \; \; \mathrm{s} \; \mathrm{cm}^{-3}
\end{equation}
and for the shocked ejecta:~
\begin{equation}\label{tau1_prime}
\hat{\tau}_2 = 1.5^{+0.6}_{-0.3} \: 10^{9} \; \xi \; \varepsilon_2^{-1/2} \; \; \mathrm{s} \; \mathrm{cm}^{-3}
\end{equation}

Now, we shall compare the ionization age given by Table \ref{res_XSPEC_2} and the mean ionization age $\hat{\tau}$ in our region both 
for the shocked ambient medium and the shocked ejecta.
The equality between these two estimates must constrain the value of $\xi \: \varepsilon^{-1/2}$, \emph{i.e.}
essentially the value of the filling factor $\varepsilon$.

For the shocked ambient medium, the lower limit on $\hat{\tau}_1$ is $3.1 \: 10^{10} \; \xi \; \varepsilon_1^{-1/2} \; \; \mathrm{s} \; \mathrm{cm}^{-3}$
(cf. Eq. \ref{tau2_prime}).
Even if $\varepsilon_1=1$ (\emph{i.e.} the shocked ambient medium fills all the volume of the projected region of A1),
we see that the mean ionization age $\hat{\tau}_1$ cannot be lower than $3.1 \: 10^{10} \; \xi$ s cm$^{-3}$.
Since $\xi \simeq 1/3$ is a reasonable choice (cf. Table \ref{filling_factor_Table} in Appendix \ref{appendix2}), we 
see that this lower limit value is inconsistent by two orders of magnitude with the value $\tau_1 = 10^{8}$ s cm$^{-3}$ given in Table \ref{res_XSPEC_2}.
This demonstrates that the model representing the shocked ambient medium is not consistent.
If the shocked ambient medium does not fill the entire volume of the projected region (\texttt{Case 2} in Appendix \ref{volume_estimation}),
it increases the lower limit on $\hat{\tau}_1$ which is already inconsistent.

Yet, the \textit{Chandra} image of the 4-6 keV continuum shows that the bulk of the X-ray emission coming
from the region A1 arises from filaments corresponding to thin shocks which extend all along the southwest. 
Then, from a morphological view point, it seems that this X-ray emission arises from a recently shocked ambient medium concentrated in a small fraction
of the entire volume of the projected region.
In that case (cf. Appendix \ref{recently_shocked}), Eqs. (\ref{xi_small}) and (\ref{tau_prime}) leads to 
$\hat{\tau}_1 = 1.6^{+1.0}_{-0.7}\: 10^{10} \; \varepsilon_1^{1/2} \; \; \mathrm{s} \; \mathrm{cm}^{-3}$.
The equality between this mean ionization age and $\tau_1$ (cf. Table \ref{res_XSPEC_2})
implies that $1.5 \: 10^{-5} \leq \varepsilon_1 \leq  2.0 \: 10^{-4}$.
We do not know any thermal model that predicts a compression of the shocked ambient medium behind the shock 
front in such a thin layer, as noted by Vink \& Laming (2003) for Cas A.
Hence, it is likely that the X-ray emission arising from this filamentary region is nonthermal in origin.

For the shocked ejecta, the equality between $\hat{\tau}_2$ (cf. Eq. \ref{tau1_prime}) and $\tau_2$ (cf. Table \ref{res_XSPEC_2}) 
implies that $\varepsilon_2 = 2.1 \: 10^{-5} \; \xi^2$. 
Thus, these ejecta fill a very small fraction of the volume of the region A1.
An upper limit (lower) on $\varepsilon_2$ is obtained by taking the maximum (minimum) values of $\hat{\tau}_2$ and the minimum (maximum) values of $\tau_2$:~
$1.4 \: 10^{-6} \; \xi^2 = \varepsilon_2^- \leq \varepsilon_2 \leq \varepsilon_2^+ = 1.1 \: 10^{-4} \; \xi^2$ at a 90\% confidence level.
These values must be compared to $\varepsilon_{2,\mathrm{Chevalier}} \simeq 0.08$
calculated from the self-similar model of Chevalier (1982) of $n=9$ (power-law index of the initial outer density profile in the ejecta),
and $s=0$ (power-law index of the initial ambient medium density profile).
In Appendix \ref{filling_factor_chevalier}, Table \ref{filling_factor_Table} indicates that other values of $n$ and $s$ do not change $\varepsilon_{2,\mathrm{Chevalier}}$ a lot.
With $\xi \simeq 1/3$, we have $\varepsilon_2 = 2.3 \: 10^{-6}$ ($\varepsilon_2^- = 1.6 \: 10^{-7}$, $\varepsilon_2^+ = 1.2 \: 10^{-5}$).
It is hard to imagine how the ejecta could be so clumpy.
With such small volumes, the density would be very high ($\hat{n}_e \sim 80 \; \mathrm{cm}^{-3}$) and
would induce short cooling timescales ($\sim 150$ years) leading to observable optical knots
in the southeastern region of Kepler's SNR
which does not seem to be the case (Bandiera \& van den Bergh 1991; Douvion et al. 2001).
Consequently, this model with two NEI components does not lead to a consistent astrophysical model for the spectrum
of the region A1.

\subsubsection{Nonthermal model\label{srcut+nei}}
We consider now models for which the emission from the shocked ambient medium is nonthermal.
The ejecta component of this model is the same as in the previous model.
As for the shocked ambient medium modelling, rather than choosing a simple power-law, 
we used a synchrotron radiation cutoff model 
(called SRCUT in XSPEC, see Reynolds \& Keohane 1999) which allows us to describe the 
synchrotron spectrum including the radio data. 
This nonthermal model embraces the following parameters:
the 1 GHz radio flux density $S_{\mathrm{1 \: GHz}}$,
the radio spectral index $\alpha$ and the rolloff frequency $\nu_r$
(\emph{i.e.} approximatively the frequency at which the flux has dropped by a factor of 10 from a straight power-law).

To estimate the local radio flux $S_{\mathrm{1 \: GHz}}$ at 1 GHz in the region A1,
we used the local radio flux at 1.35 GHz in the region A1  which is estimated to $\sim 0.13  \: \mathrm{Jy}$ from the 
radio map presented in Dickel et al. (1988).
To convert the flux at 1.35 GHz to the value at 1 GHz, we normalized it by the ratio of the total radio flux at 1 GHz of $19 \: \mathrm{Jy}$ (Green 2001)
with the total radio flux at 1.35 GHz of $15 \: \mathrm{Jy}$ (DeLaney et al. 2002), giving $S_{\mathrm{1 \: GHz}} = 0.16 \: \mathrm{Jy}$.

Then, we froze the 1 GHz radio flux at $0.16 \: \mathrm{Jy}$ in XSPEC whereas $\nu_r$ and $\alpha$ were
freely fitted.
This results in a better fit ($\chi^2 = 627$ for 602 dof) than the fit obtained with the previous model with 2 NEI ($\chi^2 = 701$ for 601 dof).
Figure \ref{synchrotron} (right panel) shows the best-fit obtained with the SRCUT+NEI model and illustrates the decomposition of the different components.
The best-fit parameters are shown in Table \ref{res_XSPEC_2}.
The fitted  value of the radio spectral index $\alpha = 0.66 \pm 0.01$
can be compared to the global spectral index of $0.64 \pm 0.02$ (Matsui et al. 1984) found for the entire SNR
and to an estimate of the local value of about $0.70 \pm 0.02$ taken from  
the spectral index radio map of Kepler's SNR made by DeLaney et al. (2002).
Our value of $\alpha$ is between these values and particularly,
the derived upper limit of our spectral index $\alpha = 0.67$ is flatter than DeLaney et al.'s value.
We check that fixing the radio flux to 0.16 and the spectral index $\alpha$ to 0.70 makes the fit worse ($\chi^2 = 1315$).
The difference between our best fit spectral index and the local (spatially and spectrally) radio index can be understood because
we measure the mean slope between the radio and the X-rays.
We note that such a curvature of the spectrum (radio index steeper than mean index) 
is predicted by the non-linear shock acceleration theory (Reynolds \& Ellison 1992).

Hence, it is likely that the emission coming from the forward shock is nonthermal in origin.
However, like in the model with two NEI components, the model representing the shocked ejecta is still not consistent.

The spectrum of the forward shock in Kepler is similar to that observed by \textit{Chandra}
in Tycho (Hwang et al. 2002) and Cas A (Vink \& Laming 2003). 
Using a different arguments, it was also argued in these papers that a significant part of the spectrum behind the forward shock was consistent with a nonthermal interpretation.
For the future, the combination of the \textit{Chandra} and \textit{XMM-Newton} data will provide a deeper understanding
and stronger constraints on the regions dominated by the forward shock emission.

The maximum electron energy $E_{max}$ can be deduced from the rolloff frequency $\nu_r$ and the
magnetic field $B$ (see Reynolds \& Keohane 1999). 
Using Table \ref{res_XSPEC_2}, we get $E_{\mathrm{max}} \left(\frac{B}{10 \: \mu \mathrm{G}}\right)^{1/2} =  65^{+12}_{-7} \: \mathrm{TeV}$.
A magnetic field of $B=14 \; \mu$G (Matsui et al. 1984) gives $E_{\mathrm{max}} =  55^{+11}_{-6}  \: \mathrm{TeV}$.
This local value is consistent with
the 50 TeV maximum electron energy found in Reynolds \& Keohane (1999) for the full SNR.

\subsection{Three components models\label{3_components}}
In the previous nonthermal model (Sect. \ref{srcut+nei}), the thermal emission from the shocked ambient medium is not modelled but it must be there.
It is possible to estimate its density by adding a third component.

We add another NEI model to this NEI+SRCUT model by fixing 
all the elemental abundances to their solar values.
Moreover, the ionization age $\tau_1$ and the normalization $Norm_1$ of the shocked ambient medium
can be computed respectively from (\ref{ne}) and from the normalization formula (\ref{nh})
as a function of the mean H density $\hat{n}_{\mathrm{H}}$.
Here, we assume that $\xi = 1/3$ and the parameter $\varepsilon_1$ is fixed to $\varepsilon_{1,\mathrm{Chevalier}} \simeq 0.3$, value chosen 
from the Chevalier (1982) self-similar profiles ($n=9, s=0$).
Once the shocked gas density $\hat{n}_{\mathrm{H}}$ is fixed, $\tau_1(\hat{n}_{\mathrm{H}})$ and $Norm_1(\hat{n}_{\mathrm{H}})$ are frozen and the data are fitted.
We have included in XSPEC an upper limit on the temperature $T_1$
given by the relation $T_s = \frac{3}{16} \: \frac{\mu}{k} \: m_{\mathrm{H}} \: v_s^2$
with $v_s = \frac{n-3}{n-s} \: \frac{r_s}{t_s} \simeq 4550 \: \mathrm{km/s}$ (cf. Appendix \ref{appendix2}) and then
we set $T_1 \leq T_s = 25$ keV.
It was found that the third component does not improve the fit.
Then the plot of the $\chi^2$ curve as a function of $\hat{n}_{\mathrm{H}}$ gives
an upper limit on $\hat{n}_{\mathrm{H}}$, \emph{i.e.} on the shocked ambient medium density.
We find $\hat{n}_{\mathrm{H}} \leq 0.15$ cm$^{-3}$ at a 90\% confidence level 
which indicates that the forward shock propagates in a very tenuous ambient medium in the southeast.

\section{Conclusion}

The \textit{XMM-Newton} observation has yielded several new results on Kepler's SNR.
Observational facts and their astrophysical interpretations are summarized here:
\begin{itemize}
\item The Si K line emission extends to a higher radius than the Fe L line emission, notably in the southern part of the remnant.
It indicates that there is no inversion of the Si and Fe layers.
\item The Fe K line emission peaks at a smaller radius than the Fe L line emission in the north. It is a clue
for a higher temperature inwards the remnant.
\item The 4-6 keV and 8-10 keV images show that there are few variations in the hardness of the continuum over the remnant,
except in a few features.
\item The asymmetry in the Fe K emission-line is not associated with any asymmetry in the Fe K equivalent width map.
Hence, the abundance variations do not cause the north-south brightness asymmetry. The strong emission
in the north may be due to overdensities in the ambient medium. 
As for the southeastern region of the remnant, it has a very low equivalent width.
The Si K maps lead to the same conclusions.
\item The spectrum of the southeastern rim indicates that the emission of the shocked ambient medium is likely to be mostly nonthermal there.
If this is true, the H density of the shocked ambient medium in that region must be lower than $0.15$ cm$^{-3}$.
\end{itemize}

\begin{acknowledgements}
The authors thank the referee R. Bandiera for helpful comments.
G. C.-C. is indebted to J. L.~Sauvageot for his great help in data reduction. 
J. P. H. acknowledges support from NASA XMM grant NAG5-9927.
The results presented here are based on observations obtained with \textit{XMM-Newton},
an ESA science mission with instruments and contributions directly funded by ESA Member States and the USA (NASA).
\end{acknowledgements}

\appendix

\section{Optimization methods for images \label{appendix1}}

\subsection{Optimization of the energy bands width\label{appendix1.1}}
The width of the energy bands has been optimized by the
following method~: each line and the surrounding continuum are
fitted by the following model (in cts/keV):~
\begin{eqnarray}\label{dn_o_de}
\frac{dN}{dE}(E)  =  P_{\beta}(E) + \underbrace{B \exp \left(-\frac{(E-E_0)^2}{2\sigma^2}\right)}_{G(E)} + \: R(E)
\end{eqnarray}
where $P_{\beta}(E) = A \: \exp (-\beta E) \; \mathrm{or} \; P_{\beta}(E) = A \: E^{-\beta}$
according to the shape of the continuum.
$R(E)$ is a reference background (Lumb et al. 2002) which has been renormalized
in the 10-12 keV band for MOS and 12-14 keV band for pn.
The coefficients $A,\beta,B,E_0,\sigma$
of the model are evaluated by minimizing the Cash statistic (Cash 1979). Thus, the width
$[E_1 ,E_2]$ of each band is determined by optimizing the signal to noise ratio:~
\begin{equation}
S/N(E_1,E_2) \propto \int_{E_1}^{E_2} G(E) dE / \left(\int_{E_1}^{E_2} \frac{dN}{dE}(E) dE \right)^{1/2}
\end{equation}

\subsection{Continuum subtraction\label{appendix1.2}}
We can take advantage of the above method to clean line-images from the continuum contribution.

Our goal is to obtain an image of the continuum ($I_c$) which lies under the emission-line in the energy band $[E_1,E_2]$ that
will be subtracted from the initial emission-line image.
For that, we use two continuum bands $[E_1^-,E_2^-]$ and $[E_1^+,E_2^+]$
located on either side of the line band.
These two bands are chosen so that the continuum is not contamined by nearby lines.
Due to the poor statistics, it is not possible to have directly the shape of the continuum in each pixel $(i,j)$.
So, we first assume that the shape of the continuum remains constant over the SNR.

Then, the underlying continuum image $I_c$ in the band $[E_1,E_2]$ will be a linear combination of
the background-subtracted continuum images $I_{c}^{-}$ and $I_{c}^{+}$ in the bands $[E_1^-,E_2^-]$ and $[E_1^+,E_2^+]$, respectively.
Formally, $\forall (i,j)$:
\begin{equation}\label{I_c}
I_{c}(i,j) = \lambda \: \alpha_{c}^{-} \: I_{c}^{-}(i,j) \: + \: (1-\lambda) \: \alpha_{c}^{+} \: I_{c}^{+}(i,j)
\end{equation}
where the normalisation coefficients ($\alpha_{c}^{-}$,$\alpha_{c}^{+}$)
and the weight $\lambda \in [0,1]$ are determined from the spectrum integrated over the entire SNR as described below.

We compute once and for all the normalisation coefficients by
$\alpha_{c}^{-} = \mathcal{N}_{\beta_0} / \mathcal{D}_{c}^{-} 
\; \mathrm{and} \; 
\alpha_{c}^{+} = \mathcal{N}_{\beta_0} / \mathcal{D}_{c}^{+}$
where $\mathcal{D}_{c}^{-}$ ($\mathcal{D}_{c}^{+}$) is the measured background-subtracted 
continuum counts in the band $[E_1^-,E_2^-]$ ($[E_1^+,E_2^+]$).
The number of counts $\mathcal{N}_{\beta}$ in the modelled continuum of slope $\beta$ 
is defined by $\int_{E_1}^{E_2} P_{\beta}(E) dE$ in the energy band [$E_1,E_2$] 
($\mathcal{N}_{\beta}^{-}$ and $\mathcal{N}_{\beta}^{+}$ in the bands $[E_1^-,E_2^-]$ and $[E_1^+,E_2^+]$, respectively).
The value $\beta_0$ is determined from the adjustement of the entire remnant spectrum in the band $[E_1^-,E_2^+]$ as described in Sect. \ref{appendix1.1}.
If the shape of the continuum is the same over the entire SNR then, whatever $\lambda$, the underlying continuum is correctly represented by $I_c$ defined in Eq. (\ref{I_c}).

However, it is possible that the shape of the continuum changes over the SNR.
Then, our goal is to choose the weight $\lambda$ to be the least sensitive to spectral variations, \emph{i.e.} to variations of $\beta$ within the model (\ref{dn_o_de}).
In each pixel $(i,j)$, we suppose that the modelled continuum has a slope $\beta(i,j)$ ($\equiv \beta$ hereafter).
Then, in each pixel, the proportions between $I_c$, $I_{c}^{-}$, $I_{c}^{+}$ should follow those between
$\mathcal{N}_{\beta}$, $\mathcal{N}_{\beta}^{-}$, $\mathcal{N}_{\beta}^{+}$.
But, from Eq. (\ref{I_c}), $I_c$ will follow:
\begin{equation}\label{Cbeta}
\mathcal{C}_{\beta} \equiv \lambda \: \alpha_{c}^{-} \: \mathcal{N}_{\beta}^{-} + (1-\lambda) \: \alpha_{c}^{+} \: \mathcal{N}_{\beta}^{+}.
\end{equation}
By construction, we have $\mathcal{C}_{\beta_0} = \mathcal{N}_{\beta_0}$ $\forall \: \lambda$, if the spectral model $P_{\beta}$ is a good representation of the data.
To ensure that this equality remains true for small spectral variations, we choose the weight $\lambda$ such that:
\begin{equation}\label{solve_lambda}
\left( d (\mathcal{C}_{\beta} - \mathcal{N}_{\beta})/d \beta \right)_{\beta=\beta_0} = 0
\end{equation}

However, the solution $\lambda$ of Eq. (\ref{solve_lambda}) may give too much weight to the continuum map
which has the lower statistics so that statistical fluctuations are amplified.
To avoid this effect, we take the value of $\lambda$ given by Eq. (\ref{solve_lambda})
provided that the variance of Eq. (\ref{I_c})
is lower than $10\%$ of the variance of the initial emission-line.
This is equivalent to:
\begin{equation}\label{variance}
\mathcal{V}_c \equiv [\lambda \alpha_{c}^{-}]^2 \: N^{-} \: + \: [(1-\lambda) \alpha_{c}^{+}]^2 \: N^{+} \leq 0.1 \: N_{\mathrm{line}}
\end{equation}
where $N^{-}$ ($N^{+}$) is the total number of counts (\emph{i.e.} with the background) in the continuum band $[E_1^-,E_2^-]$ ($[E_1^+,E_2^+]$)
and $N_{\mathrm{line}}$ the total number of counts in the line in [$E_1,E_2$].
If the procedure gives higher variance, then we choose the closest $\lambda$ compatible with the constraint on the variance
(\emph{i.e.} we give higher weight to the continuum band with better statistics).

\subsection{Equivalent width image\label{appendix1.3}}
The interpolation based method can also be used to construct EQW images with some variants.

To build an EQW image, we have divided the image in the line
(background and continuum subtracted) by the 
underlying continuum $I_c$ (background subtracted) at $E_0$.

Because of such an operation, the statistical fluctuations can be strongly amplified.
So, we must insist on the statistical aspect rather than on the spectral aspect.
In that case, the parameter $\lambda$ of the $I_c$ image used for dividing is calculated
so that the relative error $\sqrt{\mathcal{V}_c}/\mathcal{N}_{\beta_0}$ (see Sect. \ref{appendix1.2}) 
is at most of the same order as the relative error on the initial emission-line spectrum $1/\sqrt{N_{\mathrm{line}}}$.

\section{Mean ionization age computation \label{appendix2}}
Here, we present the method used for computing a mean ionization age in the aim to check the value of the ionization age obtained by the spectral fit.

\subsection{General method\label{General_method}}
We can write the mean ionization age $\hat{\tau}$ as a function of
the mean electron density $\hat{n}_e$ which is determined from the observation,
the age of the SNR $t_s$ and a parameter $\xi$ depending on the spatial distribution, 
so that 
\begin{equation}\label{tau_mean}
\hat{\tau} = \xi \: \hat{n}_e \: t_s
\end{equation}

We first calculate the mean electron density $\hat{n}_e$ as a function of the mean H density $\hat{n}_{\mathrm{H}}$ by~:
\begin{equation}\label{ne}
\hat{n}_e = (\sum_i z_{i} Ab_i Ab_{\odot,i}) \hat{n}_{\mathrm{H}}
\end{equation}
where $z_{i}$ is the number of electrons lost by the atom $i$ (fully or partly ionized),
$Ab_i$ and $Ab_{\odot,i}$ respectively the abundance with respect to solar used by XSPEC and 
the solar abundance\footnote{For solar abundances, $\sum_i z_{i} Ab_i Ab_{\odot,i} \simeq 1.21$.} of the element $i$,
$i$=\{H,He,C,N,O,Ne, Mg,Si,S,Ca,Fe,Ni\}. 
Then, from the normalization formula given in XSPEC, we can deduce the mean H density:
\begin{equation}\label{nh}
\hat{n}_{\mathrm{H}} = \left( \frac{4 \pi \: D^2}{10^{-14} \; V}\frac{Norm}{\sum_i z_{i} Ab_i Ab_{\odot,i}}\right)^{1/2} \; \; \mathrm{cm}^{-3}
\end{equation}
where $D$ is the distance to the source (in cm), 
$Norm$ the normalization value given by XSPEC and $V$ the emission volume which projects onto the local extraction region
(we shall explain below how to estimate the local volume $V$).
Hence, $\hat{n}_e$ is calculated with (\ref{ne}) using (\ref{nh}).

If the spatial structure is known (from a model), it is possible to evaluate $\xi$.
The ionization age should be estimated by averaging over emission measure (see Kaastra \& Jansen 1993)
\begin{equation}\label{U}
\hat{\tau} = [\int \: \tau(r) \: n_e(r)^2 \: r^2 \: dr] \; / \; [\int \: n_e(r)^2 \: r^2 \: dr]
\end{equation}
and the mean electron density from the emission measure
\begin{equation}\label{D}
\hat{n}_e = \left( [\int \:  n_e(r)^2 \: r^2 \: dr] \; / \; [\int \: r^2 \: dr] \right)^{1/2}.
\end{equation}
so that $\xi = \hat{\tau} / (\hat{n}_e \: t_s)$.
For reference, $\hat{\tau}$ and $\hat{n}_e$ can be computed using a self-similar hydrodynamical model (Chevalier 1982).
In Eq. (\ref{U}), $\tau(r) = \int_{t_0(r)}^{t_s} \: n_e(t) \: dt$ where $t_0(r)$ is the time 
when a mass element located at the position $r$ at $t_s$ was shock-heated by the blast wave.
In Eqs. (\ref{U}) and (\ref{D}), the integration is made over the volume of the considered shell (shocked ejecta or shocked ambient medium).
Table \ref{filling_factor_Table} gives some values of $\xi$ for different values of the
power-law index $n$ of initial outer density profile in the ejecta and the power-law index $s$ of the initial outer density profile for the ambient medium.

\begin{table}[t]
\centering
\begin{tabular}{rrcccc}
\hline
    &     & \multicolumn{2}{c}{Shocked ambient medium} & \multicolumn{2}{c}{Shocked ejecta} \\
$n$ & $s$ & $\xi_1$ & $\varepsilon_{1,\mathrm{Chevalier}}$ & $\xi_2$ & $\varepsilon_{2,\mathrm{Chevalier}}$ \\ \hline \hline
 7 & 0 & 0.34 & 0.39 & 0.35 & 0.11 \\
 9 & 0 & 0.31 & 0.32 & 0.32 & 0.08 \\ 
12 & 0 & 0.29 & 0.29 & 0.30 &  0.05 \\ \hline
 7 & 2 & - & 0.54 & - &  0.04 \\
 9 & 2 & - & 0.49 & - &  0.03 \\
12 & 2 & - & 0.46 & - &  0.02 \\ \hline
 7 & -1 & 0.28 & 0.34 & 0.28 &  0.14 \\
 9 & -1 & 0.23 & 0.28 & 0.24 &  0.10 \\
12 & -1 & 0.21 & 0.25 & 0.21 &  0.06 \\ \hline
\end{tabular}
\caption{Normalized emission-measure averaged ionization parameters ($\xi_1$ and $\xi_2$)
and filling factors ($\varepsilon_{1,\mathrm{Chevalier}}$ and $\varepsilon_{2,\mathrm{Chevalier}}$) 
in the self-similar model of Chevalier (1982) for different values of $n$ and $s$.
$n$ is the power-law index of the initial outer density profile in the ejecta
and $s$ the power-law index of the initial ambient medium density profile.
$s=0$ corresponds to an homogeneous ambient medium, $s=2$ to an ambient medium modified by a wind and
$s=-1$ simulates a dense shell in the ambient medium.
The parameters $\xi_1$ and $\xi_2$ are calculated using Eqs. (\ref{U}) and (\ref{D}).
It is not possible to give the value of $\xi_1$ and $\xi_2$ for $s=2$ because in that case
both ionization age and density profiles increase to infinity as they approach the contact discontinuity.}
\label{filling_factor_Table}
\end{table}

\subsection{Volume estimation\label{volume_estimation}}
Here, we intend to describe how we estimate the emission volume $V$ which projects onto the local extraction region.
This estimation depends on the local geometry of the SNR but can be treated with a global model depending on the situation encountered.

Two cases should be considered:
\begin{description}
\item[1,] The first one corresponds to the situation in which the projected region is contained
between the forward shock and the reverse shock radii (respectively $r_s$ and $r_r$) of the projected shell of the entire SNR.
\item[2,] The second one corresponds to the situation in which the shocked ejecta and the shocked ambient medium
partially fill the volume $V$ which projects onto the local extraction region.
\end{description}

\paragraph{Case 1:}
The volume $V$ that we integrate along the line-of-sight in the local region is always a cylinder full of
shocked ejecta and shocked ambient medium whatever is the position of the reverse shock. 
So the fact that we use local or global parameters to estimate the volume which projects 
onto the local extraction region does not matter.

The volume $V$ can be computed as a function of the overall volume of the SNR considered as a filled sphere
whose radius $r_s = D \: \theta_s$ is given by the position of the forward shock
\footnote{For Kepler's SNR, we set $\theta_s = 120 \arcsec$ 
from the 4-6 keV continuum image by choosing the same coordinates center as in Hughes (1999).
This angular radius corresponds to $r_s = 2.8 \pm 0.8$ pc.}.
We define $\alpha$ so that $V = \alpha  \: \frac{4}{3} \pi \: (D \: \theta_s)^3$.
Afterwards, we correct $V$ by a filling factor $\varepsilon \leq 1$
which takes into account the volume fraction filled by the shocked gas (ambient or ejecta).

In this configuration, the mean ionization age using Eqs. (\ref{tau_mean}), (\ref{ne}) and (\ref{nh}) becomes:~
\begin{eqnarray}\label{tau_prime}
\hat{\tau} = 9.2 \: 10^{11} \; \left(\frac{t_s}{1 \; \mathrm{yr}}\right) \; \left(\frac{\theta_s}{1 \arcsec}\right)^{-3/2} \; \left(\frac{D}{1 \; \mathrm{kpc}}\right)^{-1/2} \nonumber\\
\times \left(Norm \; \sum_i z_{i} Ab_i Ab_{\odot,i}\right)^{1/2} \: \frac{\xi}{\left(\varepsilon \: \alpha\right)^{1/2}} \; \; \mathrm{s} \: \mathrm{cm}^{-3}
\end{eqnarray}

\paragraph{Case 2:\label{filling_factor_chevalier}}
In that case, Eq. (\ref{tau_prime}) is still valid but we apply a filling factor $\varepsilon$ which takes into account the average properties of the SNR.

Table \ref{filling_factor_Table} gives some values of filling factors calculated in the framework of the 
self-similar model of Chevalier (1982) averaged over the remnant.
Let $r_s$ be the radius of the forward shock, $r_r$ the radius of the reverse shock radius and
$r_c$ the radius of the contact discontinuity.
The filling factor of the shocked ambient medium is given by the geometry of the shell and is equal
to $\varepsilon_{1,\mathrm{Chevalier}} = (r_s^3 - r_c^3) / r_s^3$.
On the other side, the filling factor of the shocked ejecta is equal to $\varepsilon_{2,\mathrm{Chevalier}} = (r_c^3 - r_r^3)/r_s^3$.

\subsection{Recently shocked gas\label{recently_shocked}}
The thermal fits (cf. Sect. \ref{2 nei}) result in very small ionization age $\tau$.
This could happen if the X-ray gas was very recently shocked.

We assume that the speed of the shock $v_s$ is constant on the time scale $\xi \: t_s$ where $t_s$ is the SNR age.
Let $n_0$ be the mean ambient density and $\Omega$ the compression ratio ($\Omega > 4$).
The amount of shocked ambient medium matter in the region A1 is $\Omega \: n_0 \: \varepsilon \: V$ where
$V$ is the volume that we integrate along the line-of-sight in the considered region and $\varepsilon$
a filling factor (see Sect. \ref{volume_estimation}). 
It is also $S \: v_s \: \xi \: t_s \: n_0$ where
$S$ is the surface encountered by the shock in the region A1.
By taking an upper limit $S_{\mathrm{max}}$ on the surface shocked by the blast wave, we obtain a
lower limit $\xi_{\mathrm{min}}$ on $\xi$.
With $S_{\mathrm{max}} = 4 \: r_{\mathrm{A1}} \: r_s$ where $r_{\mathrm{A1}} = D \: \theta_{\mathrm{A1}}$ is the radius of the projected region A1
and $r_s$ the shock radius, we obtain:
\begin{eqnarray}
\xi_{\mathrm{min}}  & = & 4.97 \: 10^3 \; \left(\frac{D}{1 \; \mathrm{kpc}}\right) \; \left(\frac{\theta_s}{1 \arcsec}\right)^{2} \; \left(\frac{\theta_{\mathrm{A1}}}{1 \arcsec}\right)^{-1} \nonumber \\
& &  \times \; \left(\frac{t_s}{1 \; \mathrm{yr}}\right)^{-1} \; \left(\frac{v_s}{1 \; \mathrm{km/s}}\right)^{-1} \; \alpha \; \Omega \; \varepsilon \label{xi_small}
\end{eqnarray}
This equation shows that $\xi$ is geometrically related to the filling factor $\varepsilon$ 
(cf. Sect. \ref{volume_estimation})\footnote{Using Kepler's SNR parameters in the region A1
$D = 4.8 \pm 1.4 \; \mathrm{kpc}$, $\theta_s = 120 \arcsec$, $\theta_{\mathrm{A1}} = 14 \arcsec$, $t_s = 400 \; \mathrm{yr}$,
$v_s = 4550 \; \mathrm{km/s}$, $\alpha = 0.008$, $\Omega = 4$ leads to the lower limit
$\xi_{\mathrm{min}} \simeq 0.43 \: \varepsilon \pm 0.13$.}.
A small value of $\xi$ implies a very small
filling factor especially as we took a rough upper limit on the surface $S$.



\end{document}